\documentclass[fleqn,usenatbib]{mnras}

\usepackage{newtxtext,newtxmath}
\usepackage[T1]{fontenc}
\usepackage{float}
\DeclareRobustCommand{\VAN}[3]{#2}
\let\VANthebibliography\thebibliography
\def\thebibliography{\DeclareRobustCommand{\VAN}[3]{##3}\VANthebibliography}

\usepackage{graphicx, subcaption}	% Including figure files
\usepackage{amsmath}	% Advanced maths commands
\usepackage{xcolor}
\usepackage{threeparttable, tablefootnote}
\title[Subaru-HSC survey of Cygnus OB2]{Subaru Hyper Suprime-Cam Survey of Cygnus OB2 Complex - I: Introduction, Photometry and Source Catalog}

\author[Gupta et al. ]{Saumya Gupta$^{1}$\thanks{kcsaumya.gupta@gmail.com}, 
Jessy Jose$^{1}$\thanks{jessyvjose1@gmail.com}, 
Surhud More$^2$,
Swagat R. Das$^{1}$,
Gregory J. Herczeg$^{3}$,
\newauthor
Manash R. Samal$^{4}$,
Zhen Guo$^{5}$,
Prem Prakash$^{1}$,
Belinda Damian$^{6}$,
Michihiro Takami$^{7}$,
\newauthor
Satoko Takahashi$^{8,9}$,
Katsuo Ogura$^{10}$,
Tsuyoshi Terai$^{11}$,
Tae-Soo Pyo$^{11,12}$
\\
$^{1}$Indian Institute of Science Education and Research (IISER) Tirupati, Rami Reddy Nagar, Karakambadi Road, Mangalam (P.O.), Tirupati 517 507, India\\
$^{2}$Inter University Centre for Astronomy and Astrophysics, Ganeshkhind, Pune 411007, India\\
$^{3}$ Kavli Institute for Astronomy and Astrophysics, Peking University, Yi He Yuan Lu 5, Haidian Qu, Beijing 100871, China\\
$^{4}$Physical Research Laboratory (PRL), Navrangpura, Ahmedabad 380 009, Gujarat, India\\
$^{5}$ Centre for Astrophysics Research, University of Hertfordshire, Hatfield AL10 9AB, UK\\
$^{6}$ Christ ({\it Deemed to be University}), Bangalore, India \\
$^{7}$ Institute of Astronomy and Astrophysics, Academia Sinica 11F of Astronomy-Mathematics Building, National Taiwan University, Taiwan, R.O.C\\
$^{8}$ Joint ALMA Observatory, Alonso de C\'ordova 3107, Vitacura, Santiago, Chile\\
$^{9}$ NAOJ Chile, National Astronomical Observatory of Japan, Alonso de C\'ordova 3788, Office 61B, Vitacura, Santiago, Chile, 7630492\\
$^{10}$ Kokugakuin University, Higashi, Shibuya-ku, Tokyo 150-8440, Japan\\
$^{11}$ Subaru Telescope, National Astronomical Observatory of Japan, National Institutes of Natural Sciences, 650 North Aohoku Place Hilo, HI 96720, USA\\
$^{12}$ School of Mathematical and Physical Science, SOKENDAI (The Graduate University for Advanced Studies), Hayama, Kanagawa 240-0193, Japan\\
}

\date{Accepted XXX. Received YYY; in original form ZZZ}

\pubyear{2021}

\begin{document}
\label{firstpage}
\pagerange{\pageref{firstpage}--\pageref{lastpage}}
\maketitle

\begin{abstract}
    Low mass star formation inside massive clusters is crucial to understand the effect of cluster environment on processes like circumstellar disk evolution, planet and brown dwarf formation. The young massive association of Cygnus OB2, with a strong feedback from massive stars, is an ideal target to study the effect of extreme environmental conditions on its extensive low-mass population.%The extreme cluster environment of the young massive association of Cygnus OB2 makes it a unique potential test-bed to verify the role of stellar feedback from surrounding massive stars on the evolution of low mass and sub-stellar mass objects in the region.
    %The extreme cluster environment and the abundance of both massive and low-mass stellar population, thus makes the young massive association of Cygnus OB2 a unique target for our study
    We aim to perform deep multi-wavelength studies to understand the role of stellar feedback on the IMF, brown dwarf fraction and circumstellar disk properties in the region. We introduce here, the deepest and widest optical photometry of 1.5$^\circ$ diameter region centred at Cygnus OB2 in r$_{2}$, i$_{2}$, z and Y-filters using %Thanks to the rare combination of high sensitivity and wide coverage of
    Subaru Hyper Suprime-Cam (HSC). This work presents the data reduction, source catalog generation, data quality checks and preliminary results about the pre-main sequence sources. We obtain 713,529 sources in total, with detection down to $\sim$ 28 mag, 27 mag, 25.5 mag and 24.5 mag in r$_{2}$, i$_{2}$, z and Y-band respectively, which is $\sim$ 3 - 5 mag deeper than the existing Pan-STARRS and GTC/OSIRIS photometry. %The presented HSC photometry reaches $\sim$ 3 - 5 mag deeper than the existing Pan-STARRS and GTC/OSIRIS photometry for the region. 
    We confirm the presence of a distinct pre-main sequence branch by statistical field subtraction of the central 18$^\prime$ region. We find the median age of the region as $\sim$ 5 $\pm$ 2 Myrs with an average disk fraction of $\sim$ 9$\%$. At this age, combined with A$_V$ $\sim$ 6 - 8 mag, we detect sources down to a mass range $\sim$ 0.01 - 0.17 M$_\odot$. The deep HSC catalog will serve as the groundwork for further studies on this prominent active young cluster.
   %We confirm the presence of a distinct pre-main sequence branch, occupied by $\sim$ 88$\%$ of the previously identified young stellar objects (YSOs) and estimate the cluster's age to be $<$ 10 Myrs. The obtained HSC catalog will serve as the groundwork for further studies on this massive outside the solar neighborhood.

%We present here the deepest and the widest optical photometry of 1.5 degree diameter region centred at Cygnus OB2, a young massive Galactic star forming region outside the solar neighborhood, using the Subaru HSC in 4 filters $r_{2}$, $i_{2}$, z and Y. The source catalog consists of a total of 713,529 sources with detection down to $26.8 mag$ in $i_{2}$-band which corresponds to $\sim$ 0.037 M$_\odot$ i.e $<$ Lithium-burning objects. A distinct pre-main sequence branch is observed in the region and by isochrone fitting, the age of the cluster is estimated to be $<$ 10 Myrs in accordance with the literature. We further aim to perform a multi-wavelength study of the region to identify and classify YSOs in the region and quantify the role of external photo-evaporation on Circumstellar disk evolution, star/Brown dwarf fraction and behavior of IMF in the low mass regime of Cygnus OB2.
\end{abstract}


%%%%%%%%%%%%%%%%%%%%%%%%%%%%%%%%%%%%%%%%%%%%%%%%%%

%%%%%%%%%%%%%%%%% BODY OF PAPER %%%%%%%%%%%%%%%%%%

\section{Introduction}
The complete stellar life cycle is significantly shaped by its mass, which is in-turn determined by the less understood evolutionary stages of star formation and its related processes (\citealt{doi:10.1146/annurev-astro-081811-125528, 2015arXiv150906382A,article} and references therein). 
As low-mass stars ($<$ 1-2 M$_\odot$) spend comparatively longer time in the rudimentary stages than their massive counterparts ($>$ 8 M$_\odot$), comprehensive studies on low-mass star formation can provide useful insight into the interesting underlying processes like protoplanetary disk formation and evolution (\citealt{Hartmann_2008, 2011ARA&A..49...67W, 2015arXiv150906382A}), %the fundamental behaviour of IMF in the low mass sub-stellar regime
brown dwarf formation and the factors affecting them (\citealt{basu2017perspectives, 2019BAAS...51c.333M}). Moreover, since most of the stars form in clusters, hence cluster environment plays a crucial role in stellar evolution and related processes (\citealt{refId0, 2015A&A...581A...5S, Jose_2016, 2021MNRAS.502.2665P, 2021MNRAS.504.2557D}). For example, disk evolution has been observed to be affected by various factors like viscous accretion (\citealt{2015ApJ...804...29G, 2017RSOS....470114E}), stellar density (\citealt{2018MNRAS.478.2700W}), external photoevaporation in diverse harsh environments like ONC (\citealt{1993ApJ...410..696O}), NGC 1977 (\citealt{2016ApJ...826L..15K}), Cygnus OB2 (\citealt{2012ApJ...746L..21W, 2016arXiv160501773G, 2019MNRAS.485.1489W}). Another intriguing question which requires further investigation is the ambiguous uniformity of Initial Mass Function (IMF) and its behavior in the low-mass and sub-stellar regime. Although many recent and past studies suggest a uniform IMF across various star forming regions in the Milky Way (\citealt{2010ARA&A..48..339B,2014prpl.conf...53O, 2016EAS....80...73M, 2017ApJ...836...98J, 2021MNRAS.504.2557D}), variation has been observed in the extreme environments like the Galactic Center (e.g. \citealt{Lu_2013,  2019ApJ...870...44H}), least luminous Milky Way satellites (\citealt{2013ApJ...771...29G, 2018ApJ...855...20G}) and massive elliptical galaxies (\citealt{2010Natur.468..940V, 2012Natur.484..485C}).  %Although various power laws (Salpeter 1955; Kroupa 2001; Chabrier 2003 and De Marchi 2005, 2010) describe the IMF behavior quite well in high-mass realm, they deviate from each other significantly in low-mass end. 

Since, both Galactic and extragalactic star formation principally occurs in clusters and OB-associations (e.g \citealt{2000AJ....120.3139C, doi:10.1146/annurev.astro.41.011802.094844, 2012A&A...545A.122P}), an empirical model for low mass star formation developed by eclectic inferences drawn from both Galactic as well as extragalactic studies, is a pre-requisite to answer these fundamental questions. %However, it is not possible to study the intricate details of star formation, especially the low mass end, in extragalactic realm due to observational constraints. 
However, due to observational constraints with the current technology, we can only start by analysing the relatively distant young massive Galactic star forming regions using powerful observing facilities. % Most of the Galactic studies exploring low mass end are insufficient too due to being shallow for distant regions (LMC, SMC) or focussed mainly on nearby regions in solar neighbourhood (for e.g Gould Belt regions (Dunham et al. 2015)). 
The nearby clusters (for e.g Gould Belt regions), which are the focus of most of the studies (\citealt{Dunham_2015, Dzib_2018, 2020AstBu..75..267B, 2021arXiv210205589K, 2021MNRAS.504.2557D}) are not the representative samples of extragalactic star-forming regions, where most of the star formation occurs in the extreme cluster environments of giant molecular complexes. The deep and wide field surveys of distant young massive Galactic clusters are need of the hour as such clusters are less dynamically evolved and hence, provide a robust sample of stars with similar history of formation in extreme environments (e.g. \citealt{doi:10.1146/annurev-astro-081309-130834, 2014prpl.conf..291L}). The primary goal of this work is to obtain good quality deep observations and use them to carry out an elaborate study of Cygnus OB2, a young massive Galactic cluster with extreme environmental conditions analogous to that of extragalactic star forming regions.\\
%which may be applicable on a global scale for both Galactic and extragalactic environments.\\

Cygnus OB2 ({\it Right Ascension}: 20:33:15, {\it Declination}: +41:18:54), located at $\sim$ 1.6 kpc (\citealt{10.1093/mnras/stz2548}) from the Sun, is a typical analogue of the extragalactic massive star forming regions located outside the solar neighborhood. It is the central massive OB-association  ( 2 -- 10 $\times$ 10$^4$ M$_\odot$ as determined by \citet{2000A&A...360..539K, 2010ApJ...713..871W}) embedded in the giant Cygnus X molecular complex (\citealt{2006A&A...458..855S, 2008hsf1.book...36R}) and %is justifiably termed as a proto-globular cluster \citep{2000A&A...360..539K} since it 
harbors $\sim$ 220 OB-type stars (\citealt{2012A&A...543A.101C, 2020A&A...642A.168B}) along with tens of thousands of low mass stars (\citealt{2007A&A...464..211A, 2008MNRAS.386.1761D, 2009ApJS..184...84W}). The OB2 association has an estimated age of $\sim$ 3 -- 5 Myrs (\citealt{2008MNRAS.386.1761D, 2010ApJ...713..871W, 10.1093/mnras/stv323}) and is affected by variable extinction, A$_V$ ranging between $\sim$ 4 - 8 mag \citep{10.1093/mnras/stv323}. With a cluster environment impinged by high energy radiation from massive OB-stars in the association, Cygnus OB2 is an ideal laboratory to study the role of stellar feedback on the surrounding low-mass stellar population in the region. The presence of globules and proplyds (see Figure \ref{fig:Proplyds} in Appendix \ref{sec:proplyds} for HSC r$_{2}$-band images of the known proplyds from \citet{2012ApJ...746L..21W}) in the surrounding region (\citealt{2012A&A...542L..18S, 2012ApJ...746L..21W, 2016A&A...591A..40S}) and a reduced circumstellar disk fraction in the vicinity of massive O-type stars \citep{2016arXiv160501773G} suggest the effect of ongoing external photoevaporation on disk evolution. Approximately 1843 candidate young stellar objects (YSOs) have been identified based on their NIR excess properties (\citealt{Guarcello_2013}) within an area $\sim$ 1$^{^\circ}$ $\times$ 1$^{^\circ}$ of Cygnus OB2. The GTC-OSIRIS optical study by \citet{2012ApJS..202...19G} covers the central 40$^\prime$ $\times$ 40$^\prime$ region of the Cygnus OB2 with photometry of the sources reaching down to $\sim$ 25 mag in r'-band, however photometric error exceeds 0.1 mag for $\sim$ 40$\%$ of the total sources in the catalog. Similarly, previous studies regarding the kinematics, structure as well as mass function of Cygnus OB2 are confined to a stellar mass of $\sim$ $>$ 1 M$_\odot$ (\citealt{2010ApJ...713..871W, 10.1093/mnras/stv323, 2012A&A...543A.101C, 2020MNRAS.495.3474A}). However, the low mass regime of the region covered by $<$ 0.5 M$_\odot$ stars, remains unexplored. Cygnus OB2 is thus, a potential young massive cluster for which deep and wide-field optical and NIR studies are essential. This paper is a step towards a detailed study of one of the most massive star forming regions outside the solar neighbourhood with detections reaching down to the sub-stellar regime ($\le$ 0.07 M$_\odot$).\\

We present here the deepest (r$_{2} \sim$ 28 mag) and the widest (1.5${^\circ}$ diameter) (see Figures \ref{fig: CygOB2}) optical catalog of one of the most massive Galactic star forming regions i.e Cygnus OB2 along with the preliminary analysis for a limited area using the presented HSC data. Thanks to the superb wide-field imaging capabilities of Subaru Hyper Suprime-Cam (HSC), we have obtained high quality deep optical photometry which is useful to give an insight into the low mass star formation, proto-planetary disk evolution and the effect of feedback from massive stars on the cluster properties like Initial Mass Function (IMF), star formation efficiency and star to brown dwarf ratio. \\

%The hierarchical star formation as suggested by the spatial sub-structuring of stars in the region (Wright et al. 2014; Berlanas et al. 2019) and multiple age stellar population (Wright \& Drake 2009; Wright et al. 2010) in the region remain an open area of further investigation its low mass star population. This paper is a step towards a detailed study of one of the most massive star forming regions outside the solar neighbourhood with detections reaching down to very faint limit ($\le$ 0.06 M$_\odot$) i.e sub-stellar regime, can reveal the effects of cluster environment on circumstellar disk evolution, sub-stellar formation, the nature of IMF which can further be extrapolated to extragalactic environments. \\ %The primary goal of this study is to carry out a comprehensive deep study of the Galactic analogues of massive extragalactic star forming regions. Such a study using state-of-the-art observing capabilities is highly essential to design low mass stellar models which may be applicable on a global scale for both Galactic and extragalactic environments. \\

This paper is divided into the following sections: The Section \ref{sec:data} interprets the Subaru Hyper Suprime-Cam observations, data reduction and catalog generation using HSC pipeline. Section \ref{sec:quality} presents the data quality in terms of photometry, astrometry, completeness of the HSC data along with comparison relative to already available optical photometry. In  Section \ref{sec:analysis} we present the data analysis and results obtained, aided with color-magnitude diagrams, age analysis and disk fraction analysis. We then discuss and interpret the results obtained with this data so far in Section \ref{sec:discuss} and encapsulate the entire work along with our future plans, finally in Section \ref{sec: sumup}.\\

\section{Observations and Data Reduction}
\label{sec:data}

\subsection{HSC Observations}
\begin{figure*}
	%\centering
	\includegraphics[scale = 0.6]{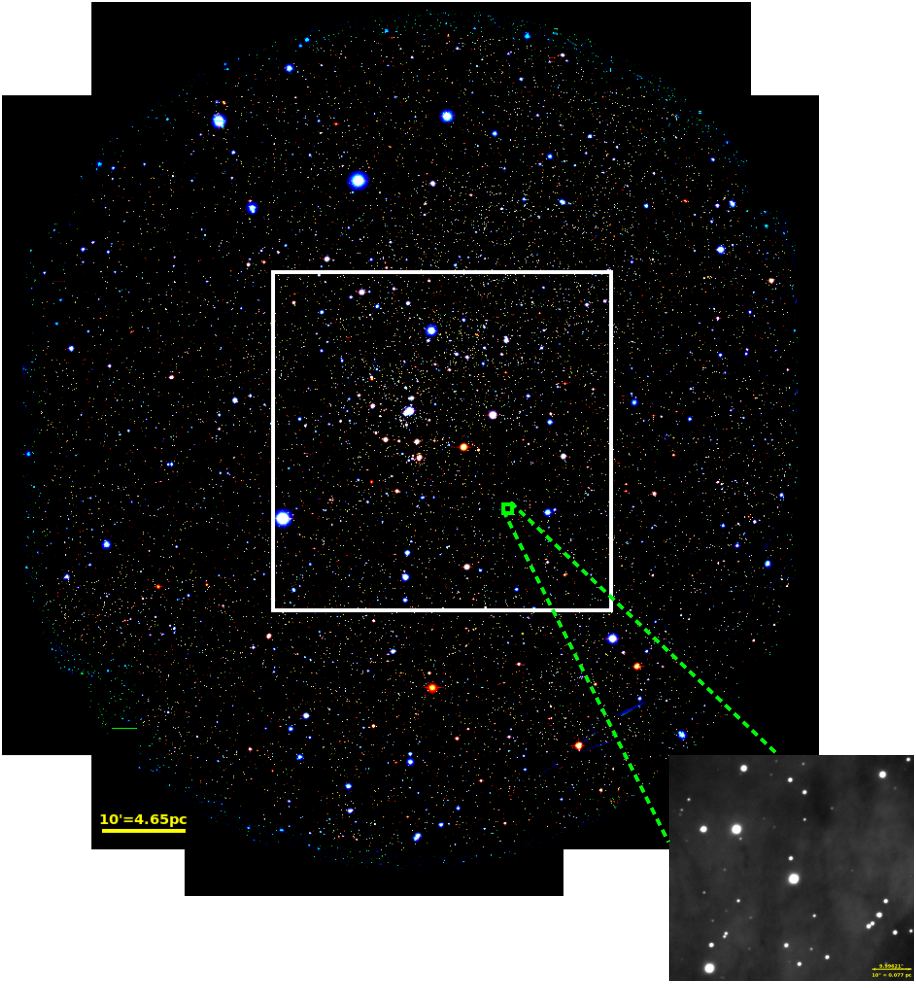}
	%\includegraphics[scale = 0.2]{cyg_centre1.eps}
	%\includegraphics[scale = 0.38]{r2-Y_vs_r2.eps}%\scriptsize
	%\includegraphics[scale = 0.4]{z-K_vs_z.eps}%\scriptsize
	%\begin{small}  
	%\scriptsize
	%\linespread{0.8}
	\caption{RGB image of the  1.5 $^\circ$ diameter region centred at Cygnus OB2 (RA: 20:33:15; Dec: +41:18:54) obtained with r$_{2}$, i$_{2}$ and Y-bands of Subaru HSC. The inset white box covers the 40$^\prime$ $\times$ 40$^\prime$ (18.6 pc $\times$ 18.6 pc) region observed by the past GTC/OSIRIS observations (\citealt{Guarcello_2013}). The inset green box covers 1$^\prime$ $\times$ 1$^\prime$ region ((RA: 20:32:12.7220; Dec: +41:06:58.778)), further zoomed in the right corner of the image which gives a vivid view of the abundance and high resolution of the point stellar sources achieved by our observations of the target region.
	 }
%{\it Top left:}
% }
%\end{small}
\label{fig: CygOB2}
\end{figure*}

\begin{figure*}
	%\centering
	%\includegraphics[scale=0.2]{FWHM_hist.png}
	%\includegraphics[scale=0.12]{FWHM_spatial.png}
	\includegraphics[width=9.2cm, height=5.7cm]{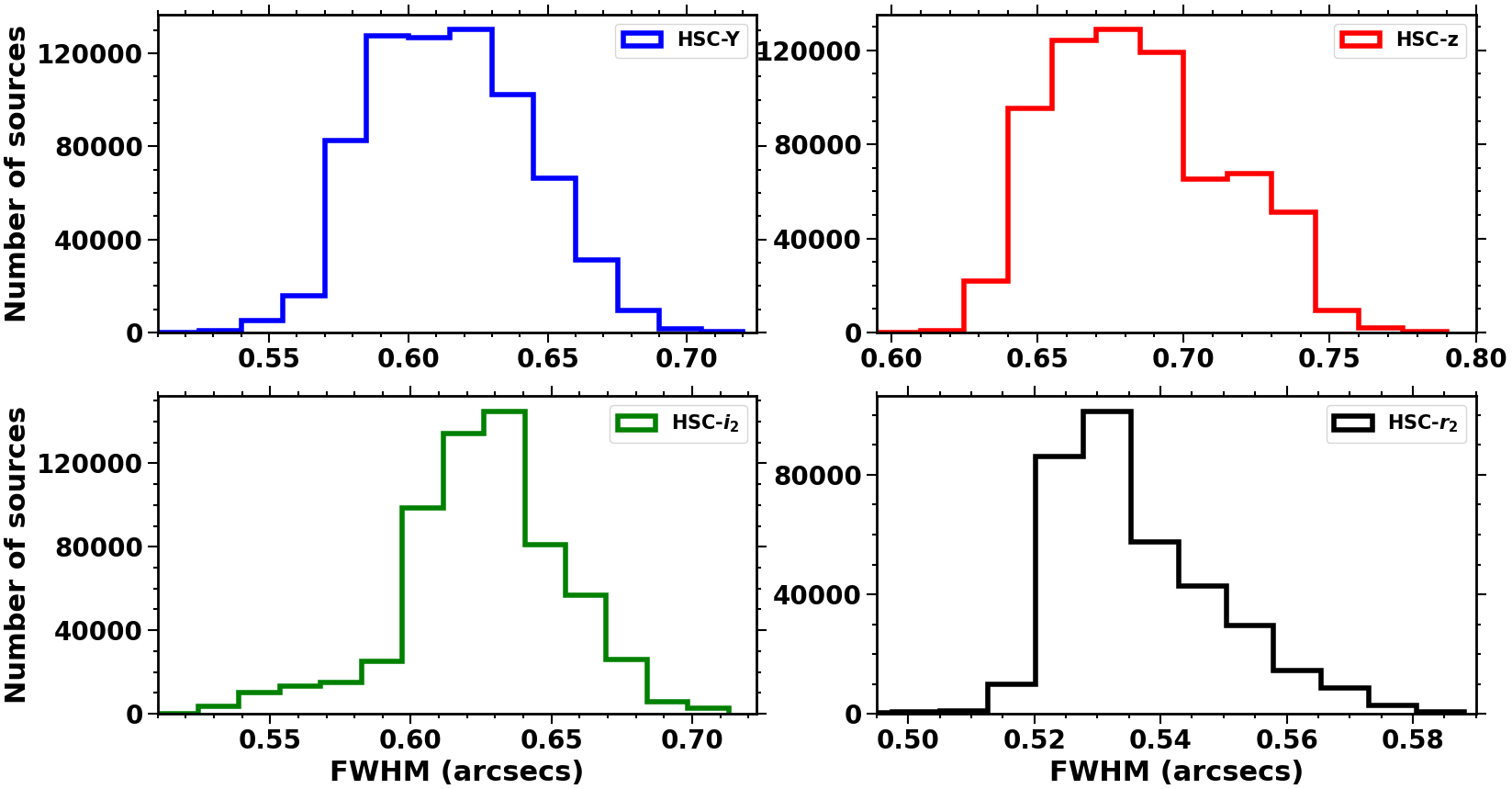}
	\includegraphics[width=8.3cm, height=5.2cm]{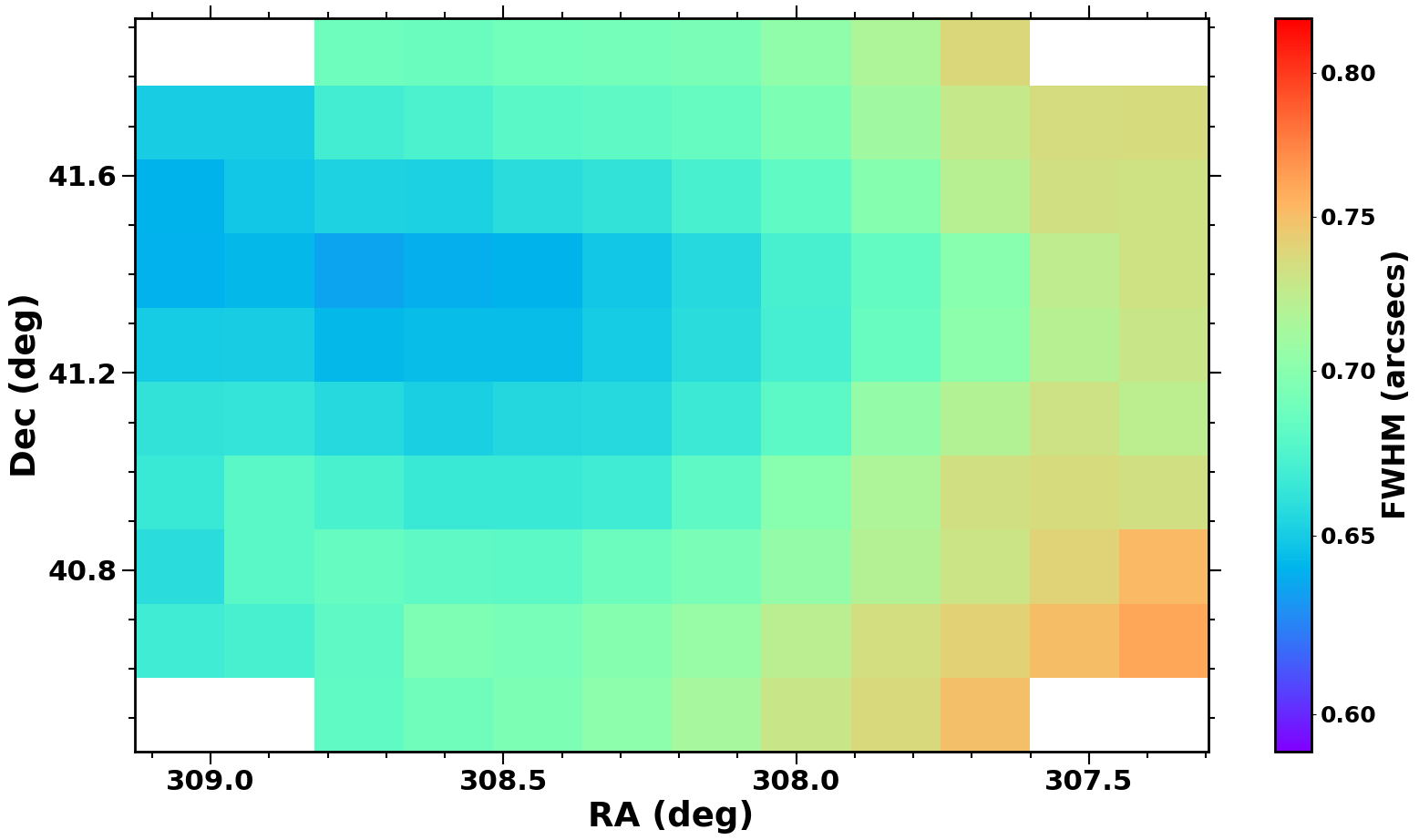}
	%\includegraphics[width=9cm, height=5.7cm]{r2-Y_vs_r2.eps}%\scriptsize
	%\includegraphics[width=8.2cm, height=5.7cm]{z-K_vs_z.eps}%\scriptsize
	%\begin{small}  
	%\scriptsize
	%\linespread{0.8}
	\caption{{\it Left}: Histogram distribution for FWHM in each HSC-filter i.e Y, z, i$_2$ and r$_2$. {\it Right}: Spatial distribution map of FWHM in z-band for the observed region. The spatial map is obtained by binning the RA and Dec parameter space into 10$^{\prime}$ $\times$ 10$^{\prime}$ bins across the entire observed region. The colorbar indicates the mean FWHM of each bin.
	 }
%{\it Top left:}
% }
%\end{small}
\label{fig:fwhm}
\end{figure*}

Subaru is an 8.2 m class optical-infrared telescope built and operated by the National Astronomical Observatory of Japan (NAOJ). With an 870 Megapixels mosaic CCD camera comprising of 116 2k $\times$ 4k CCDs with a pixel scale $\sim$ 0.17$^{\prime\prime}$, the Hyper Suprime-Cam (HSC) instrument installed at the prime focus of the telescope provides an excellent image quality over a wide field of view (FOV; 1.8 $\deg^2$) (\citealt{2012SPIE.8446E..0ZM, 10.1093/pasj/psx069, 10.1093/pasj/psx079, 2018PASJ...70S...1M}). %With 870 Megapixels camera and a pixel scale of $\sim$ 0.17", it has a wide coverage with a field of view of 1.8 $\deg^2$, 1.5 $\deg$ in diameter. 
We observed a region of 1.5$^\circ$ diameter centered at Cygnus OB2 (see Figure \ref{fig: CygOB2}) with Subaru HSC in 4 broad-band optical filters, namely, r$_{2}$, i$_{2}$, z and Y (\citealt{2018PASJ...70...66K}) on 17th September'2017 (PI: J.Jose; Program ID: S17B0108N), using EAO (East Asian Observatory) time\footnote{This EAO time for Cygnus OB2 observations was a compensatory time given to us for the ToO event GW170817, which happened during our scheduled night}. Several long exposure and short exposure frames (details given in Table \ref{tab: HSC Observation Specifications}) were taken to enhance the photometric accuracy of both faint as well as bright stars. The excellent seeing conditions ($\sim$ 0.5$^{\prime\prime}$ - 0.7$^{\prime\prime}$) atop Mauna Kea during the observations (1.07 $\le$ airmass $\le$ 1.35) and superb optics of the camera with a focal length $\sim$ 18320 mm have effectively enabled the otherwise difficult pairing of a wide field of view with detailed spatial resolution (see Figure \ref{fig: CygOB2}). The mean FWHM values achieved in individual HSC filters are indicated in Table \ref{tab: HSC Observation Specifications} and Figure \ref{fig:fwhm} {\it Left}. The achieved FWHM in individual filters is approximately uniform across the observed FOV (Figure \ref{fig:fwhm} {\it Right}). \\%, have played a crucial role in obtaining good quality images (see Figure \ref{sec:rgb} in Appendix \ref{sec:rgb}) detection of very faint sub-stellar mass objects (see Section \ref{sec:completeness}).\\

Hitherto, HSC has primarily been used for extra-galactic observations (e.g. \citealt{2019ApJ...883..183M, Ishikawa_2020, 10.1093/mnras/staa1062}). However, there is a dire lack of similar observations in Galactic stellar fields with HSC. This study is a pioneering work to utilize the powerful and highly sensitive imaging capabilities of Subaru Hyper Suprime-Cam for observations of young Galactic star forming regions. A summary of the various procedures followed and the modifications introduced in the default pipeline parameters to reduce the observed HSC data is presented below.  \\
\begin{table*}
	\centering
	\caption{ Details about short and long exposure frames and FWHM in individual filters.% Here, Exposure Time$_{short}$: Short exposure time which is given in the format (Exposure time of each frame in seconds) $\times$ (Number of short exposure frames) in individual filter. Same format applies for Exposure Time$_{long}$.
	}
	\begin{tabular}{|l|l|l|l|l|} % four columns, alignment for each
		\hline
		Filters & HSC-Y & HSC-z & HSC-i$_{2}$ & HSC-r$_{2}$ \\
		\hline
		Exposure Time$_{short}$ & 30s $\times$ 5 frames & 25s $\times$ 3 frames & 25s $\times$ 3 frames & 30s $\times$ 3 frames \\
		Exposure Time$_{long}$ & 200s $\times$ 3 frames & 300s $\times$ 4 frames & 300s $\times$ 10 frames & 300s $\times$ 16 frames \\
		Mean FWHM & 0.61$^{\prime\prime}$ & 0.68$^{\prime\prime}$ & 0.62$^{\prime\prime}$ & 0.53$^{\prime\prime}$ \\
		\hline
	%\caption{Table 1 shows various parameters obtained by analysis of final HSC catalog in individual filters. Here}
	\label{tab: HSC Observation Specifications}

	\end{tabular}
\end{table*}

\subsection{Data Reduction and Catalog Generation}
\label{sec: data reduction}

The observed raw data was downloaded from STARS (Subaru Telescope Archive System) and reduced with the help of HSC Pipeline version 6.7. The entire process of the data reduction by HSC pipeline (\textsc{hscPipe}) can be broadly classified into (1) Single-visit Processing (2) Joint Calibration (3) Coaddition (4) Coadd Processing/ Multiband Analysis. %However, a few prior steps like making of Bias, Dark, Flat, Fringe and Sky data are required for detrending and calibrating each CCD data. \\
For details regarding the following processes, refer to \citet{10.1093/pasj/psx080, 10.1093/pasj/psx081,2019PASJ...71..114A}. \\

The \textsc{hscPipe} initiates the data reduction with single-visit processing. The detrending of the raw data includes overscan subtraction, bias correction, dark current subtraction, flat-fielding, and fringe subtraction. The \textsc{hscPipe} then performs Instrument Signature Removal (ISR) to mask and interpolate the defects such as bad pixels, cross-talk, and saturated pixels. A few bright sources short-listed using a 50$\sigma$ threshold value are used as reference to model the Point Spread Function (PSF) using PSFEx software. The astrometric and photometric calibration of these sources is performed with respect to the Pan-STARRS DR1 PV3 reference catalog using the `Pessimistic B` matching algorithm\footnote{refer {\it https://dmtn-031.lsst.io/\#pessimism-of-the-algorithm} for details}. %The Pessimistic B algorithm yields precise astrometric solutions for the detected sources even in crowded fields.
We discard the default 'Optimistic B' algorithm as it is well-suited for low density fields like extragalactic realms and has failure modes in comparatively high density Galactic regions\footnote{See https://dmtn-031.lsst.io/}, which results in false matches and incorrect astrometry of the detected sources. %Hence, when implemented for the calibration of our observed crowded field data by the HSC pipeline, this algorithm generates false matches and incorrect astrometry for the detected sources. 
After performing sky subtraction and source measurements\footnote{The source measurement step includes centroiding, shape measurements, aperture corrections, etc.}, the previously generated PSF model is used to generate a deeper catalog of stars using a 5$\sigma$ threshold. The above explained process including the source extraction using 5$\sigma$ detection threshold, is performed for each CCD during single visit processing. An internal calibration is then carried out across different observing shots, termed as visits. The astrometric and photometric calibrations are carried out by matching the wcs and flux scale of each visit with the previously generated matched list of reference bright stars and corresponding corrections are then applied to each visit. \\

In the next step, the \textsc{hscPipe} coadds the images from various visits to create a single deeper image and a PSF model is constructed for the coadded image. The sky correction applied prior to the coadd process is turned off as it contaminates our coadded images due to high amount of nebulosity present in the region. The sky correction applied at this step merely writes a new background model without modifying the photometry of detected sources. We coadd the long exposure visits and short exposure visits separately for individual filters, to obtain precise photometry for some of the bright sources which get saturated in the long exposure images. Eventually, \textsc{hscPipe} performs multiband analysis in order to generate the final photometric catalog for each band. The source extraction is performed again, this time on the coadded images using 5$\sigma$ threshold value to detect sources and photometry is subsequently performed on the coadded images in each filter. As a result of the source extraction, certain above-threshold regions called footprints are generated each of which, comprises of one or more discrete sources. These footprints, containing several peaks are merged together across different filters. The overlapping footprints from different filters are then combined. Within each of such combined footprints, the peaks close enough to each other (that is, lying within 0.3$^{\prime\prime}$ of the nearest peak) are merged as one peak, otherwise are assigned as an individual new peak. This results in consistent peaks and hence, footprints across individual filters. Each of the peak corresponds to individual objects. The peaks within individual footprints are further deblended in individual filters and the total flux is apportioned into them.\\

The number of stellar sources detected during image coaddition relies upon the footprint size as each footprint consists of several blended individual peaks. The larger the size of the footprint, the more peaks or distinct objects it can hold. As the \textsc{hscPipe} is designed primarily for sparse regions, the default footprint size defined by the pipeline i.e 10$^6$ pixels is insufficient to detect all stellar point sources in a comparatively dense Galactic star forming region like Cygnus OB2. Hence, after performing rigorous checks with several footprint sizes, we finally increased it to 10$^{10}$ pixels for i$_{2}$, z and Y filters. The footprint size is increased to 10$^{11}$ pixels for r$_{2}$ filter however, to ensure maximum detection inspite of it's high sensitivity to the extensive nebulosity present in the region. %Since even with a footprint size of 10$^{10}$ pixels many of the point sources remained still undetected, we increased the footprint size further to 10$^{11}$ pixels.
The modified footprint sizes in individual filters aid in yielding an exhaustive catalog of point sources to be detected in the images. Finally, \textsc{hscPipe} performs source measurements and photometry for the detected sources and thus, % 4 different types of photometry i.e PSF photometry, Aperture photometry, CModel photometry and Kron photometry\footnote{Please refer to Bosch et al. 2018; Aihara et al. 2019 for the detailed explanation of each type of photometry.} are carried out for all the detected sources by HSC pipeline. Since PSF photometry is the most suitable for point source flux measurements and has higher S/N ratio, we adopt PSF fluxes to verify the photometric accuracy of our source catalog and performing further analysis. 
both long exposure and short exposure catalogs are obtained in r$_{2}$, i$_{2}$, z and Y bands. However, these catalogs in individual filters are contaminated with plenty of spurious\footnote{detections with no visible source present} detections. Hence, we have applied certain flags and external constraints to eradicate such spurious detections, which we explain in the following section. \\

\subsection{Point Source Selection}
\label{sec:source select}

\begin{table*}
	\centering
	\begin{threeparttable}
	\caption{ shows various flags applied with their description.}
	\label{tab:Flags}
	\begin{tabular}{ l l } % four columns, alignment for each
		\hline
		Flagging Condition & Description \\
		\hline
		deblend\_nchild != 0 & Objects which consist of multiple child\tnote{a} peaks \\
		deblend\_skipped & Objects which contain multiple blended peaks \\
		base\_PixelFlags\_flag\_crCenter & Objects overlaps cosmic ray contaminated pixels \\
		base\_PixelFlags\_flag\_suspectCenter & Object overlaps a pixel with unreliable linearity correction \\
		base\_PixelFlags\_flag\_saturatedCenter & Object with saturated pixels \\
		base\_PixelFlags\_flag\_interpolatedCenter & Object with interpolated pixels from surrounding pixels \\
		\hline
	%\caption{Table 1 shows various parameters obtained by analysis of final HSC catalog in individual filters. Here}
	%\label{Table 1:HSC Analysis}

	\end{tabular}
	\begin{tablenotes}\footnotesize
    \item [a] Each individual source peak obtained after deblending each footprint
    \end{tablenotes}

	\end{threeparttable}
\end{table*}

We apply certain flags (see Table \ref{tab:Flags}) and external constraints to remove the spurious contamination from the obtained long and short exposure catalogs (Section \ref{sec: data reduction})  with minimal loss of genuine point sources in individual filters. % The data reduction process (Section \ref{sec: data reduction}) provides us with long exposure and short exposure catalogs separately in individual filters. However, since the obtained catalogs are contaminated with plenty of spurious detections, we have applied certain flags (see Table \ref{tab:Flags}) and external constraints to eradicate them with minimal loss of genuine point sources. 
%Specifically, we flag the sources whose central pixels are contaminated due to saturation from nearby stars, or due to cosmic ray hits, or due to the presence of rejected pixels with unreliable photometry marked as suspect, or those that had to be interpolated from neighbouring pixels. We also flag sources that are blended and only use those peaks which cannot be deblended further. 
For more details on catalog flags, please refer to \citet{10.1093/pasj/psx080}. Additionally, we select sources with photometric error $\le$ 0.1 mag in individual bands for both long and short exposure catalogs. We impose an additional constraint of internal astrometric error $\le$ 0.1$^{\prime\prime}$ to remove spurious sources without any loss of good point sources. This error in astrometry of a source is with respect to its peak PSF position in different visits (For more details please refer to Section \ref{sec:astrometry}). %After a meticulous inspection with various threshold values ranging from 0.08"-0.5" using ds9 software (Joye $\&$ Mandel 2003)
We consider only those sources which have detection in at least two filters. % This implies that all the sources in a band (say z-band) will at least have counterpart in one or more other band ($r_{2}$, $i_{2}$ or Y-band).
Since the seeing conditions during our observations varied between 0.5$^{\prime\prime}$-- 0.7$^{\prime\prime}$, we have chosen the upper limit of seeing i.e 0.7$^{\prime\prime}$, as the maximum matching radius and best match as the match selection criteria to cross match the sources between any two bands using TOPCAT tool\footnote{http://www.star.bris.ac.uk/~mbt/topcat}, in order to avoid loss of any genuine counterparts (\citealt{2018ApJS..235...36M, 2020PASJ...72...64M}). The cross-matching radius, even if reduced (e.g 0.5$^{\prime\prime}$) or increased (e.g 0.8$^{\prime\prime}$ or 1$^{\prime\prime}$) varies the census of genuine sources atmost by a few hundreds, which is a negligible amount when compared to the total number of detected sources. Similary, the specified constraints of 0.1 mag in photometric error and 0.1$^{\prime\prime}$ in the internal astrometric error have been chosen after checking and discarding several values ranging between 0.07 mag -- 0.2 mag (in photometric error) and 0.08$^{\prime\prime}$ -- 0.5$^{\prime\prime}$ (in astrometric error) as it results either in loss of numerous faint point sources or an increment in spurious detection by 5--10$\%$.\\

The availability of both short exposure and long exposure photometry for the sources has enabled us to deal with the saturated sources effectively. We consider long exposure photometry in all the bands for those sources with magnitude in Y-band $>$ 18 mag. In a similar fashion, sources with Y $\le$ 18 mag are incorporated from short-exposure catalog in all the filters. However, in addition to this, we also include short exposure photometry for sources with 18 mag $\le$ Y $\le$ 22 mag and without any long exposure photometry available for them. This is specifically done in order to include the sources which lie close to bright stars and have been missed in the photometry from long exposure. The particular threshold of Y $\le$ 22 mag is chosen after verifying that the sources with only short exposure photometry and Y $>$ 22 mag, are spurious detections and hence, discarded. This merging of short and long exposure photometry can result in missing sources or repetition of sources near the merging threshold i.e Y $=$ 18 mag and its corresponding counterparts in other filters. Hence, to deal with this we take an average of the long and short exposure magnitudes for the sources with 17.8 mag $\le$ Y $\le$ 18.2 mag and their corresponding counterparts in other filters. %Hence, to deal with this we separate out the sources with $17.8 mag \le Y \le 18.2 mag$ and its counterparts in other filters and take an average of their long and short exposure magnitudes. 
An important point to note here is that the long and short exposure photometry is merged on the basis of the threshold values 18 mag or 22 mag taken in Y-band and applied to the corresponding counterparts in other bands. %The specific threshold of Y$ = 18 mag$ in the above merging procedure has been selected after several checks with other threshold values ranging between $17 mag - 20.5 mag$. However, we find the above mentioned threshold value and the merging procedure most suitable for optimal source detection.
Finally, we perform an internal matching of the sources in the entire output catalog with the upper value of astrometric uncertainity, i.e 0.1$^{\prime\prime}$ as the matching radius to avoid any repetition of sources. Any duplicates (0.5$\%$ of the total sources) of the already detected sources in the catalog are removed in this step.\\

%An additional constraint is imposed on the astrometric error values of the detected sources to remove any leftover spurious sources without any loss of good point sources. This error in astrometry of a source is internal i.e with respect to different visits in individual bands (For more details please refer to \ref{sec:astrometry}). After a meticulous inspection with various threshold values ranging from 0.08"-0.5" using ds9 software (Joye $\&$ Mandel 2003), we have included only those sources with an astrometric error less than 0.1" in our final catalog.
%To summarise, the output catalog thus procured includes only those sources which have detection in at least any 2 filters, photometric error $\le 0.1 mag$ in all the filters and internal astrometric uncertianity $\le$ 0.1$^{\prime\prime}$. To avoid any saturation effect due to bright stars, we incorporate the short exposure photometry in all the filters ($r_{2}$, $i_{2}$, z and Y) as explained above. The key steps in this process of point source selection are briefly shown as a flowchart in Figure \ref{fig:flowchart}. \\

\begin{figure*}
	%\centering
	\includegraphics[scale = 0.5]{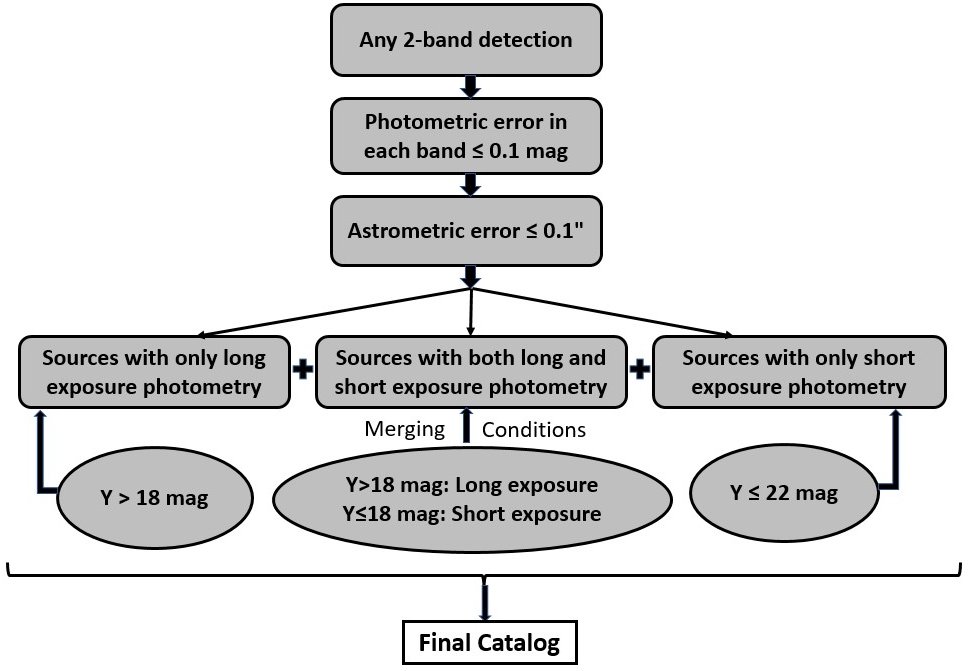}
	%\includegraphics[scale = 0.2]{cyg_centre1.eps}
	%\includegraphics[scale = 0.38]{r2-Y_vs_r2.eps}%\scriptsize
	%\includegraphics[scale = 0.4]{z-K_vs_z.eps}%\scriptsize
	%\begin{small}  
	%\scriptsize
	%\linespread{0.8}
	\caption{ This flowchart summarizes the external conditions imposed after applying flags mentioned in Table \ref{tab:Flags}. These conditions ensure the maximum point source detection and remove spurious sources from both long exposure and short exposure catalogs separately, obtained after data reduction using \textsc{hscPipe}. The short and long exposure photometry are then concatenated and merged based on conditions mentioned above. For details please refer to Section \ref{sec:source select}.
	 }
%{\it Top left:}
% }
%\end{small}
\label{fig:flowchart}
\end{figure*}

%\footnotetext[12]{Error in magnitude in individual HSC filters}

\begin{figure*}
	%\centering
	\includegraphics[scale = 0.3]{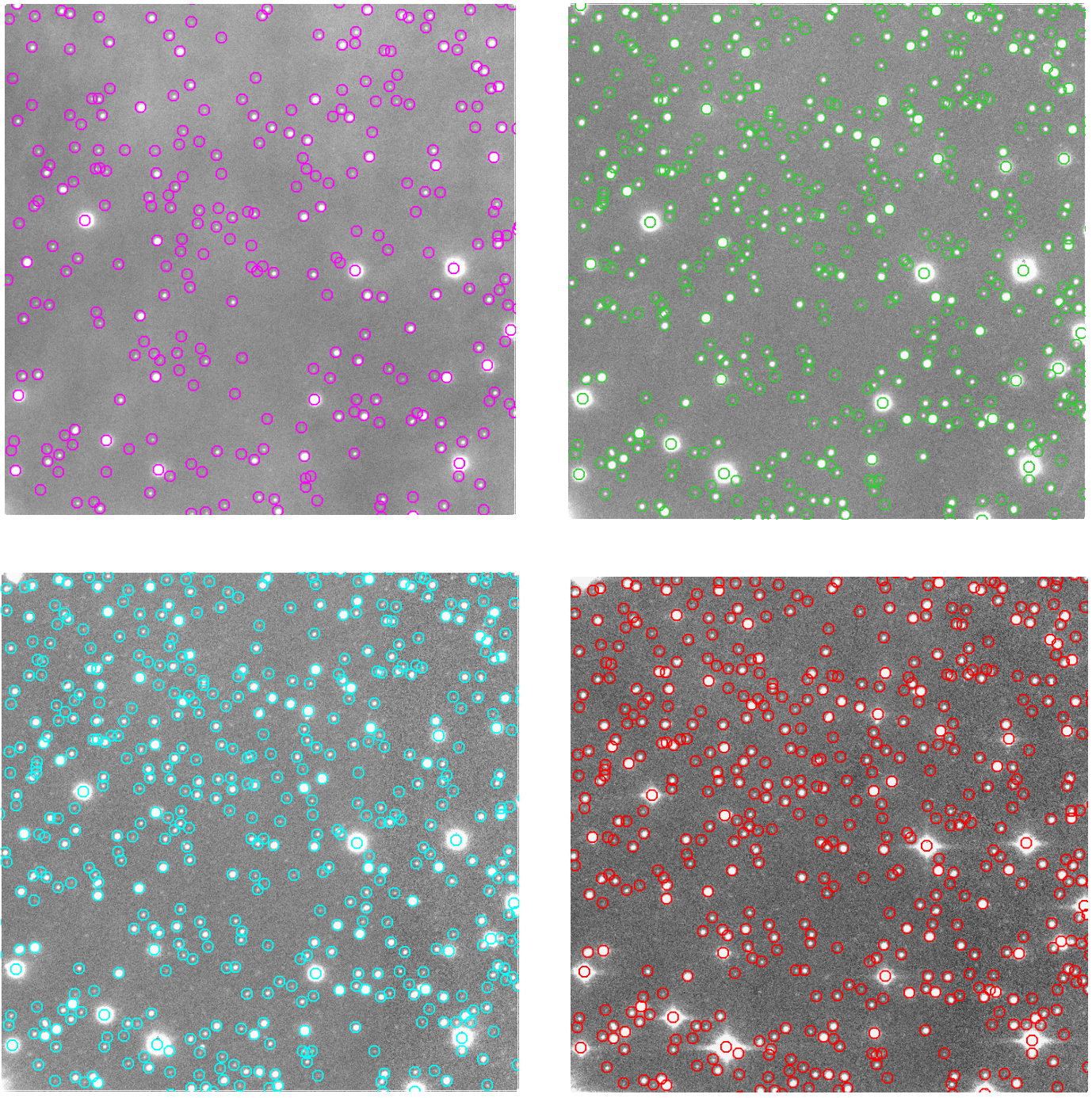}
	%\includegraphics[scale = 0.2]{cyg_centre1.eps}
	%\includegraphics[scale = 0.38]{r2-Y_vs_r2.eps}%\scriptsize
	%\includegraphics[scale = 0.4]{z-K_vs_z.eps}%\scriptsize
	%\begin{small}  
	%\scriptsize
	%\linespread{0.8}
	\caption{ A 2$^{\prime}$ $\times$ 2$^{\prime}$ area ({\it RA}: 20:34:10.4835 {\it Dec}: +40:57:48.783) in different filters is overplotted with sources detected in each individual band i.e {\it Top Left:} r$_{2}$-band, {\it Top Right:} i$_{2}$-band, {\it Bottom Left:} z-band and {\it Bottom Right:} Y-band.
	 }
%{\it Top left:}
% }
%\end{small}
\label{fig:Filter detection}
\end{figure*}

%Among several tests done for point source selection, we also attempted to select point sources based on their extendedness\footnote{difference between their CModel and PSF fluxes (Bosch et al. 2018; Aihara et al. 2019)}. Extendedness is a useful criterion for star-galaxy classification, with value = 0 for stellar point sources. However, the use of this parameter resulted in a substantial loss of genuine point sources. We suspect this to be a possible failure of the CModel measurements in such a crowded field. %{\bf If you know the galactic extinction towards this direction, one may be able to  figure out the expected galactic contamination.} %and possibly due to the presence of saturated stars. 
%We also tried to use the difference between PSF and Kron magnitudes in order to select point sources but failed to achieve any satisfactory results due to the crowded nature of our target field. Hence, we don't adopt these approaches for source selection in our data. An estimated star-galaxy classification based on Lucas et al. 2006 is presented in Appendix E.\\
To summarise, the output catalog thus procured, includes only those sources which have detection in at least any 2 filters, photometric error $\le$ 0.1 mag in all the filters and internal astrometric uncertianity $\le$ 0.1$^{\prime\prime}$. To avoid any saturation effect due to bright stars, we incorporate the short exposure photometry in all the filters (r$_{2}$, i$_{2}$, z and Y) as explained above. The key steps in this process of point source selection are briefly shown as a flowchart in Figure \ref{fig:flowchart}. We have finally secured 713,529 point sources all of which have at least a 2-band detection. Approximately, 699,798 ($\sim 98\%$) sources have Y-band photometry, 685,511 sources ($\sim$ 96$\%$) have z-band photometry, 622,011 sources ($\sim$ 90$\%$) have $i_{2}$ band photometry and 358,372 sources ($\sim$ 50$\%$) have $r_{2}$ band photometry. Figure \ref{fig:Filter detection} presents a sample of our exemplary detection in different bands for a particular region ({\it RA}: 20:34:10.4835 {\it Dec}: +40:57:48.783) of 2$^{\prime}$ $\times$ 2$^{\prime}$. Almost all the visible sources, although faint, have been successfully detected in the final HSC catalog. The adopted approach of selecting genuine point sources as described in this section has yielded the deepest and the widest comprehensive optical catalog of one of the most massive regions in the Galaxy outside the solar neighborhood. \\

%\begin{figure*}[H]
	%\centering
%	\includegraphics[scale = 0.3]{difft_det.png}
	%\includegraphics[scale = 0.2]{cyg_centre1.eps}
	%\includegraphics[scale = 0.38]{r2-Y_vs_r2.eps}%\scriptsize
	%\includegraphics[scale = 0.4]{z-K_vs_z.eps}%\scriptsize
	%\begin{small}  
	%\scriptsize
	%\linespread{0.8}
%	\caption{: {\it Left} the overplotted red circles on the individual band images signify sources detected in 3' x 3' region in $i_{2}$ and {\it Right} in Y-band after performing data reduction and shortlisting good photometry sources as described in section. 
%	 }
%{\it Top left:}
% }
%\end{small}
%\label{fig:detection}
%\end{figure*}

\section{\textbf{Data Quality}}
\label{sec:quality}

In the following sections, we discuss the data quality in terms of the photometry, astrometry, limiting magnitude of detection, completeness of the reduced HSC data with respect to the existing Pan-STARRS DR1\footnote{downloaded from https://vizier.u-strasbg.fr/viz-bin/VizieR} (\citealt{chambers2019panstarrs1}) and GTC/OSIRIS (\citealt{2012ApJS..202...19G}) optical data. We also perform a comparison of the obtained HSC photometry with respect to Pan-STARRS DR1 photometry with the help of magnitude offset plots and check the astrometric offset with respect to Pan-STARRS DR1 and Gaia EDR3 data (\citealt{2016, 2020arXiv201201533G}).

\subsection{Photometric Quality}
\label{sec:photometric quality}

\begin{figure*}
	%\centering	
	\includegraphics[scale=0.343]{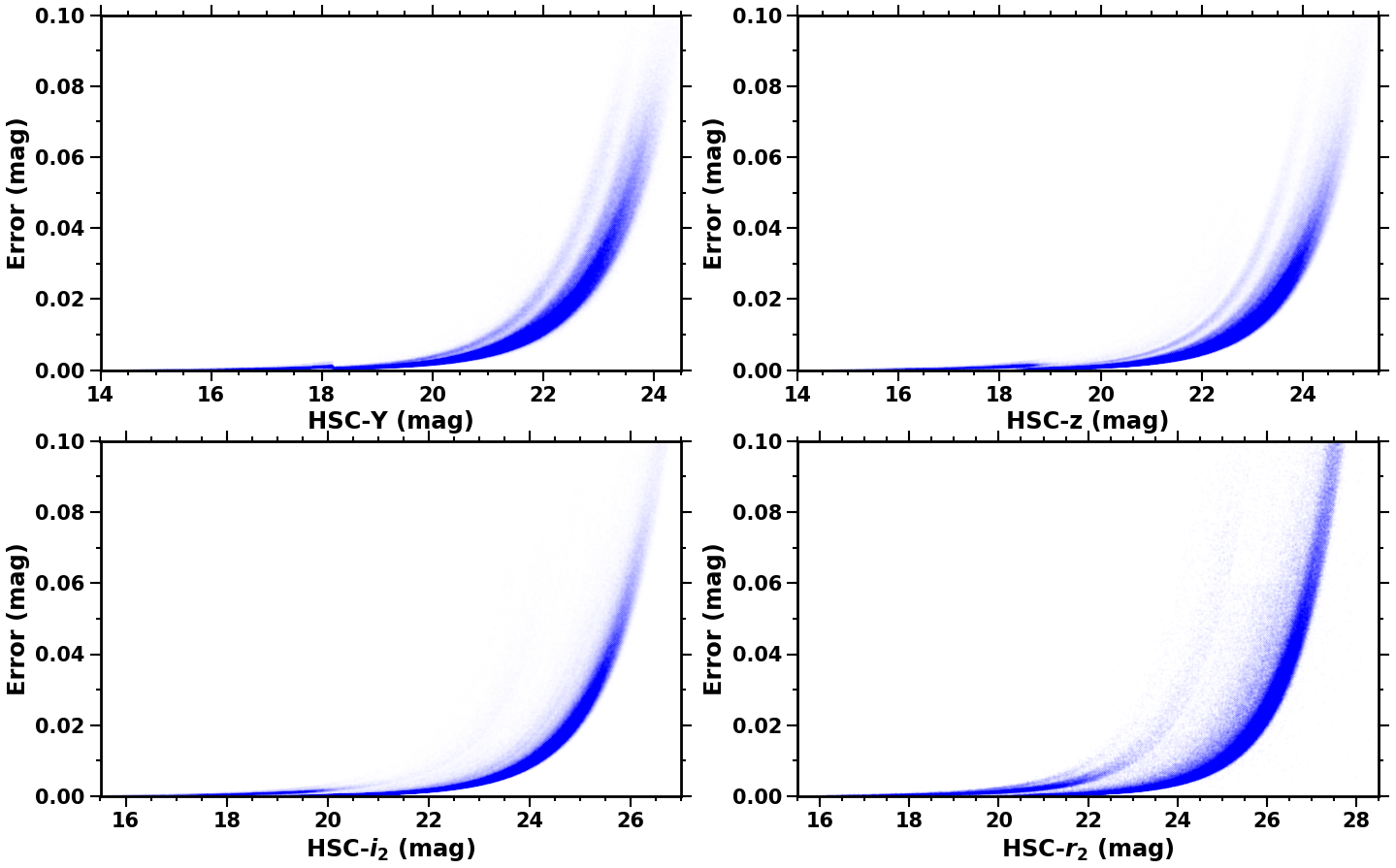}
%	\includegraphics[width=17cm, height=10cm]{Mag_err.png}
	%\includegraphics[width=11cm, height=8.7cm]{Mag_err_compare.png}
	%\includegraphics[scale = 0.2]{cyg_centre1.eps}
	%\includegraphics[scale = 0.38]{r2-Y_vs_r2.eps}%\scriptsize
	%\includegraphics[scale = 0.4]{z-K_vs_z.eps}%\scriptsize
	%\begin{small}  
	%\scriptsize
	%\linespread{0.8}
	\caption{ Scatter plots of HSC magnitudes {\it versus} error in individual HSC filters. All the sources have error $\le$ 0.1 mag. The discontinuity at Y = 18 mag in magnitude-error plot of Y-band ({\it Top Left}) is due to the merging of long and short exposure photometry. Y = 18 mag is taken as the threshold magnitude for this merging (see Section \ref{sec:source select} for details on the merging procedure.). The multiple branches observed in these plots are due to the long and short exposure photometry merged to obtain the final catalog. 
	 }
%{\it Top left:}
% }
%\end{small}
\label{fig:Mar err}
\end{figure*}

\begin{figure}
	%\centering
%	\includegraphics[width=\columnwidth, height=5.7cm]{Mag_err_compare.png}
    \includegraphics[scale=0.19]{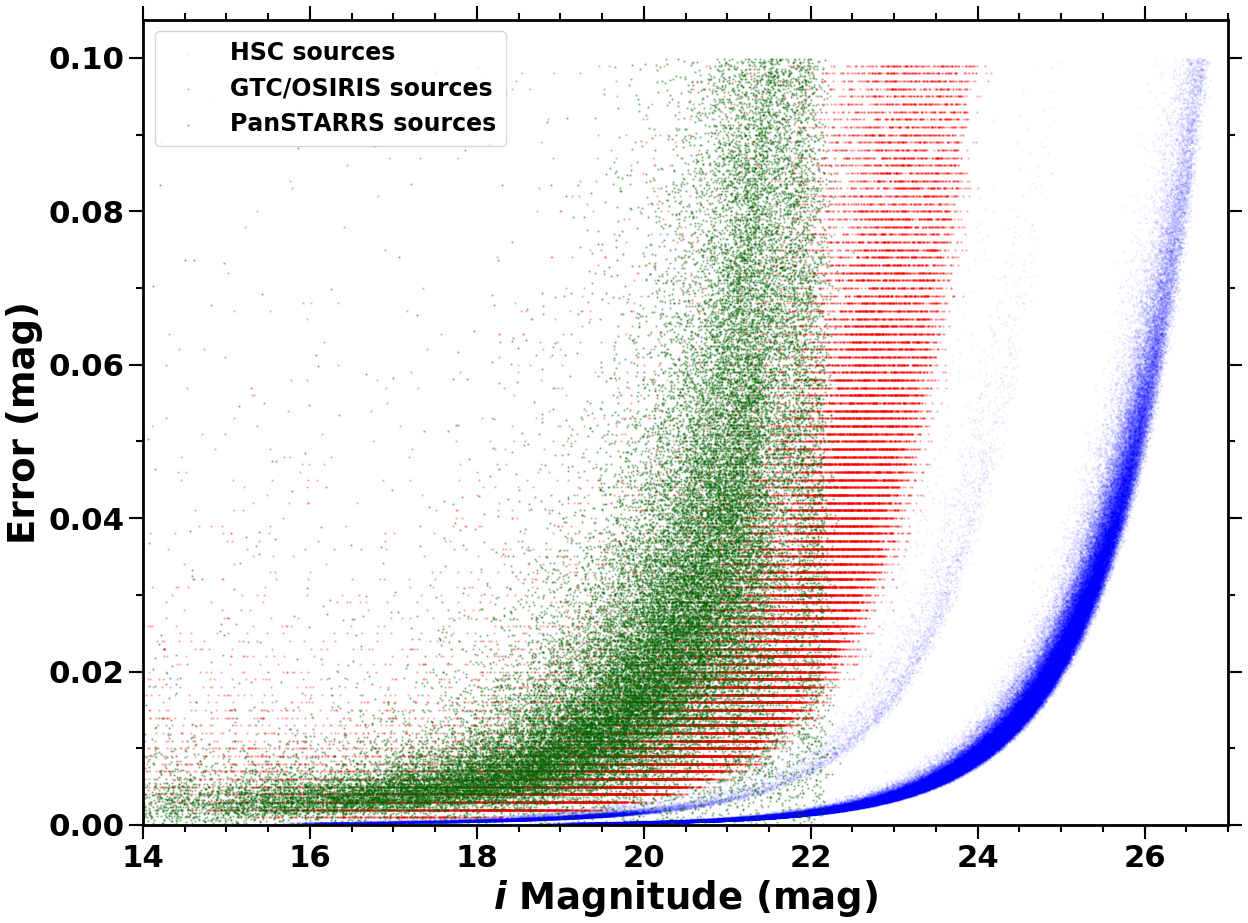}
%	\includegraphics[width=10.5cm, height=8.7cm]{Mag_err_compare.png}
	%\includegraphics[scale = 0.2]{cyg_centre1.eps}
	%\includegraphics[scale = 0.38]{r2-Y_vs_r2.eps}%\scriptsize
	%\includegraphics[scale = 0.4]{z-K_vs_z.eps}%\scriptsize
	%\begin{small}  
	%\scriptsize
	%\linespread{0.8}
	\caption{ A comparative magnitude {\it versus} error scatter plot for HSC ({\it blue}) with the existing photometry from Pan-STARRS ({\it Green}) and GTC/OSIRIS \citep{2012ApJS..202...19G} ({\it Red}) in i$_{2}$-band for an area of 30$^{\prime}$ radius centred on Cygnus OB2. The two branches observed in the HSC i$_{2}$-band plot correspond to the long and short exposure photometry.
	 }
%{\it Top left:}
% }
%\end{small}
\label{fig:Mag err compare}
\end{figure}

The error {\it versus} magnitude plots shown in Figure \ref{fig:Mar err} for the individual HSC filters i.e r$_{2}$, i$_{2}$, z and Y-filter, summarize the accuracy of the obtained HSC photometry. The plot illustrates that the photometric error is $\le$ 0.05 mag for sources with magnitudes down to $\sim$ 26.0 mag in i$_{2}$-band, 27.5 mag in r$_{2}$-band, 24.7 mag in z and 24.0 mag in Y-band. %Similarly, the detected sources have an error $<= 0.05 mag$ upto 27.4 mag in $r_{2}$-band, 24.75 mag in z and 24.0 mag in Y-band.
Approximately 91$\%$ and 95$\%$ of the total sources have a photometric error $\le$ 0.05 mag in Y and z-band respectively. Similarly, 93$\%$ of the sources detected in i$_{2}$-band and almost 90$\%$ of the detected sources in r$_{2}$-band have an error $\le$ 0.05 mag. %The multiple branches seen in the error vs magnitude plots can be attributed due to CCD noise and non- background subtraction.
A comparative error {\it versus} magnitude plot is presented in Figure \ref{fig:Mag err compare} for an area of 30$^\prime$ radius centred on Cygnus OB2 to juxtapose the accuracy of HSC photometry with previous optical studies of the region such as with Pan-STARRS, GTC/OSIRIS. Since GTC/OSIRIS observations are available for a limited FOV (40$^\prime$ $\times$ 40$^\prime$), the chosen area (30$^\prime$ radius centred at Cygnus OB2) allows a fair comparison of photometric accuracy among the HSC, Pan-STARRS and GTC/OSIRIS sources. The maximum detection limit within a photometric error $\le$ 0.1 mag attainable with Pan-STARRS and GTC/OSIRIS photometry is $\sim$ 22.5 mag--24.0 mag (i-band), which is at least 3 mag shallower when compared to the high accuracy attained by the HSC photometry down to the very faint sub-stellar regime (i$_{2} \sim$ 27.0 mag ; $\le$ 0.07 M$_\odot$) (see Section \ref{sec:completeness} and Section \ref{sec:CMD} for details). \\
\begin{figure}
	%\centering
	%\includegraphics[width=11.5cm, height=8.7cm]{Mag-off_spatial.png}
	\includegraphics[width=\columnwidth, height=5.85cm]{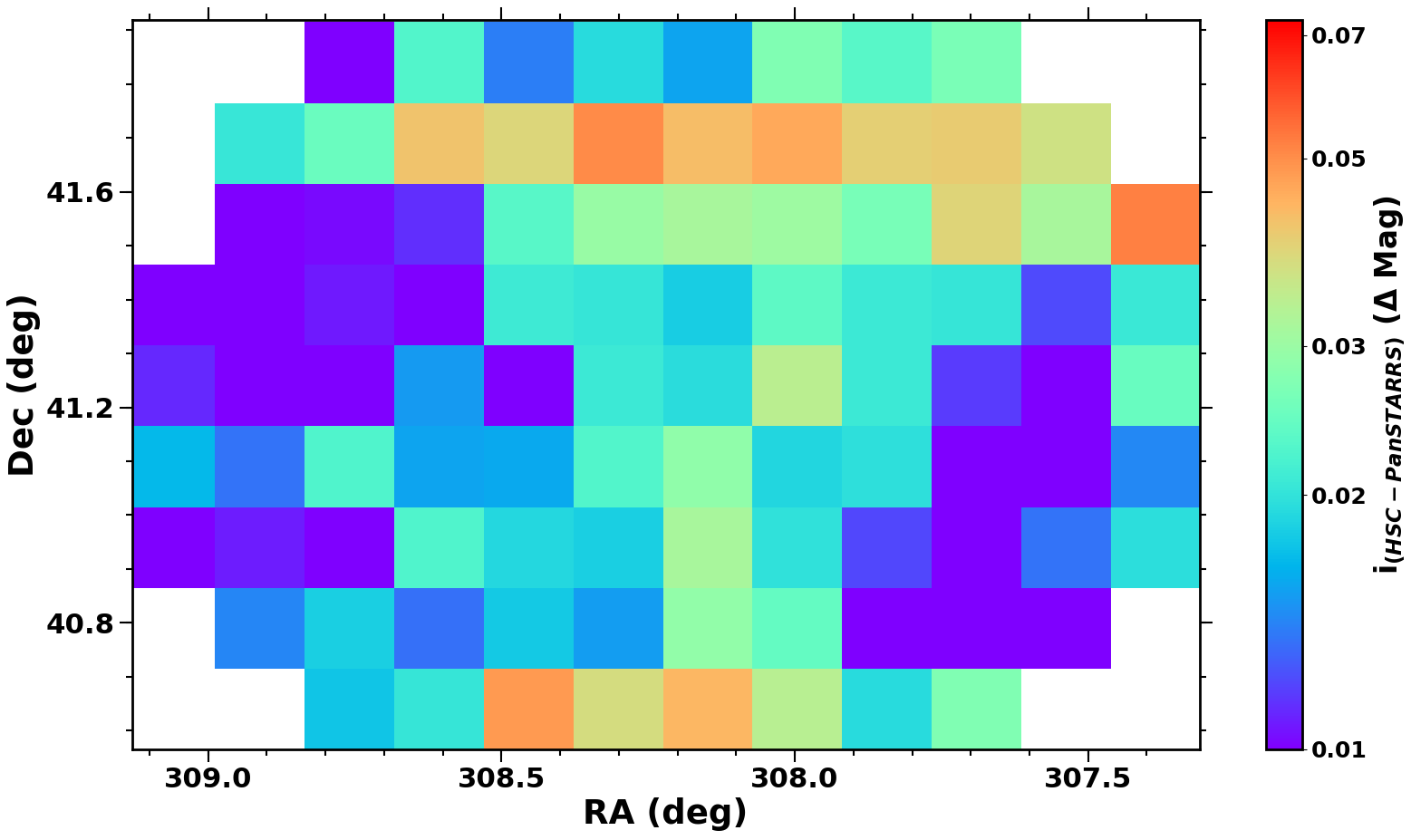}
	%\includegraphics[scale = 0.2]{cyg_centre1.eps}
	%\includegraphics[scale = 0.38]{r2-Y_vs_r2.eps}%\scriptsize
	%\includegraphics[scale = 0.4]{z-K_vs_z.eps}%\scriptsize
	%\begin{small}  
	%\scriptsize
	%\linespread{0.8}
	\caption{ Spatial distribution map generated by binning the entire observed region into 10$^\prime$ $\times$ 10$^\prime$ bins in RA and Dec parameter space to signify the variation of magnitude offset of HSC $i_{2}$-band photometry with respect to Pan-STARRS DR1 i-band photometry across the area of observations. The colorbar indicates the mean offset of sources in each bin.}
%{\it Top left:}
% }
%\end{small}
\label{fig:spatial magoff}
\end{figure}

\begin{figure*}
	%\centering
%	\includegraphics[width=14cm, height=10cm]{Mag-off_mod.png}
	\includegraphics[scale=0.343]{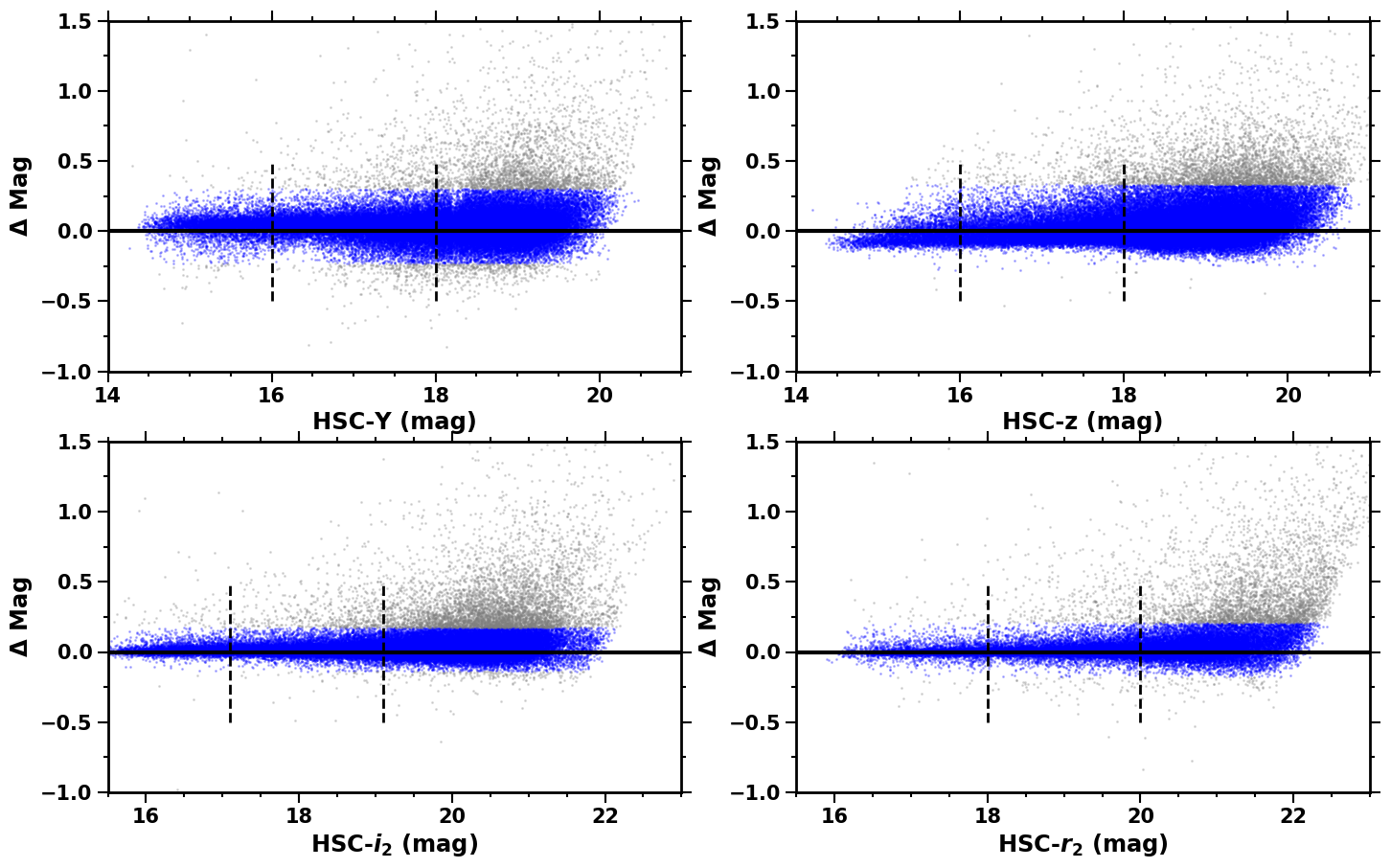}
	%\includegraphics[scale = 0.2]{cyg_centre1.eps}
	%\includegraphics[scale = 0.38]{r2-Y_vs_r2.eps}%\scriptsize
	%\includegraphics[scale = 0.4]{z-K_vs_z.eps}%\scriptsize
	%\begin{small}  
	%\scriptsize
	%\linespread{0.8}
	\caption{ Scatter plots for determining magnitude offset of HSC photometry with respect to Pan-STARRS photometry in different individual bands. An offset of 0.03$\pm0.06$ mag, 0.01$\pm0.07$ mag, 0.01$\pm0.03$ mag and 0.01$\pm0.03$ mag is observed in Y, z, $i_{2}$, $r_{2}$-band respectively for the range of magnitudes marked by dashed black lines. The marked magnitude ranges have been selected to calculate the mean magnitude offset in order to avoid the saturation of HSC photometry towards the brighter end and unreliable photometry of Pan-STARRS towards fainter end of sources. The blue sources lie within 3$\sigma$ range from mean offset whereas grey sources lie beyond 3$\sigma$ range from mean offset.
	 }
%{\it Top left:}
% }
%\end{small}
\label{fig:Magoff}
\end{figure*}
%As mentioned in Section \ref{sec: data reduction}, the photometric calibration of some reference sources was performed with respect to Pan-STARRS DR1 PV3 catalog for developing a PSF model for sources during data reduction by the HSC pipeline.
In order to assess the photometric quality, we %cross-match the HSC photometry with their Pan-STARRS counterparts using a matching radius of 1" and 
check the offset between HSC and the counterpart Pan-STARRS DR1 photometry in the individual filters. To compare the photometry, we transformed the Pan-STARRS DR1 photometry from Pan-STARRS filter system to  HSC system using the equations given in Appendix \ref{sec:transform eq}. The sources with good quality Pan-STARRS photometry have been selected by giving an error cut off  $\le$ 0.05 mag and number of stack detections $>$ 2 (\citealt{chambers2019panstarrs1}). We observe a moderate uniformity in the magnitude offset across the entire region as presented in the spatial distribution map  in Figure \ref{fig:spatial magoff}. Figure \ref{fig:Magoff} shows the scatter plots of magnitude offset i.e HSC magnitudes--Pan-STARRS magnitudes versus HSC magnitudes, in all HSC filters. A mean offset of 0.01$\pm$0.07 mag is observed in z-band with respect to the Pan-STARRS magnitudes. Similarly, other bands i.e r$_{2}$, i$_{2}$, Y-band exhibit an offset of 0.01$\pm$0.03 mag, 0.01$\pm$0.03 mag and 0.03$\pm$0.06 mag respectively, which agrees well with the offset estimated in other studies between HSC and Pan-STARRS (\citealt{Komiyama_2018, 2019PASJ...71..114A}). The mentioned mean offsets have been calculated for sources within a certain range of magnitudes (range marked by dotted black lines in Figure \ref{fig:Magoff}) in individual bands, after discarding the sources lying beyond 3$\sigma$ level (represented by grey colored dots in Figure \ref{fig:Magoff}) iteratively for 5 iterations. \\

\subsection{Astrometric Quality}
\label{sec:astrometry}

\begin{figure*}
	%\centering
	\includegraphics[width=8.6cm, height=6cm]{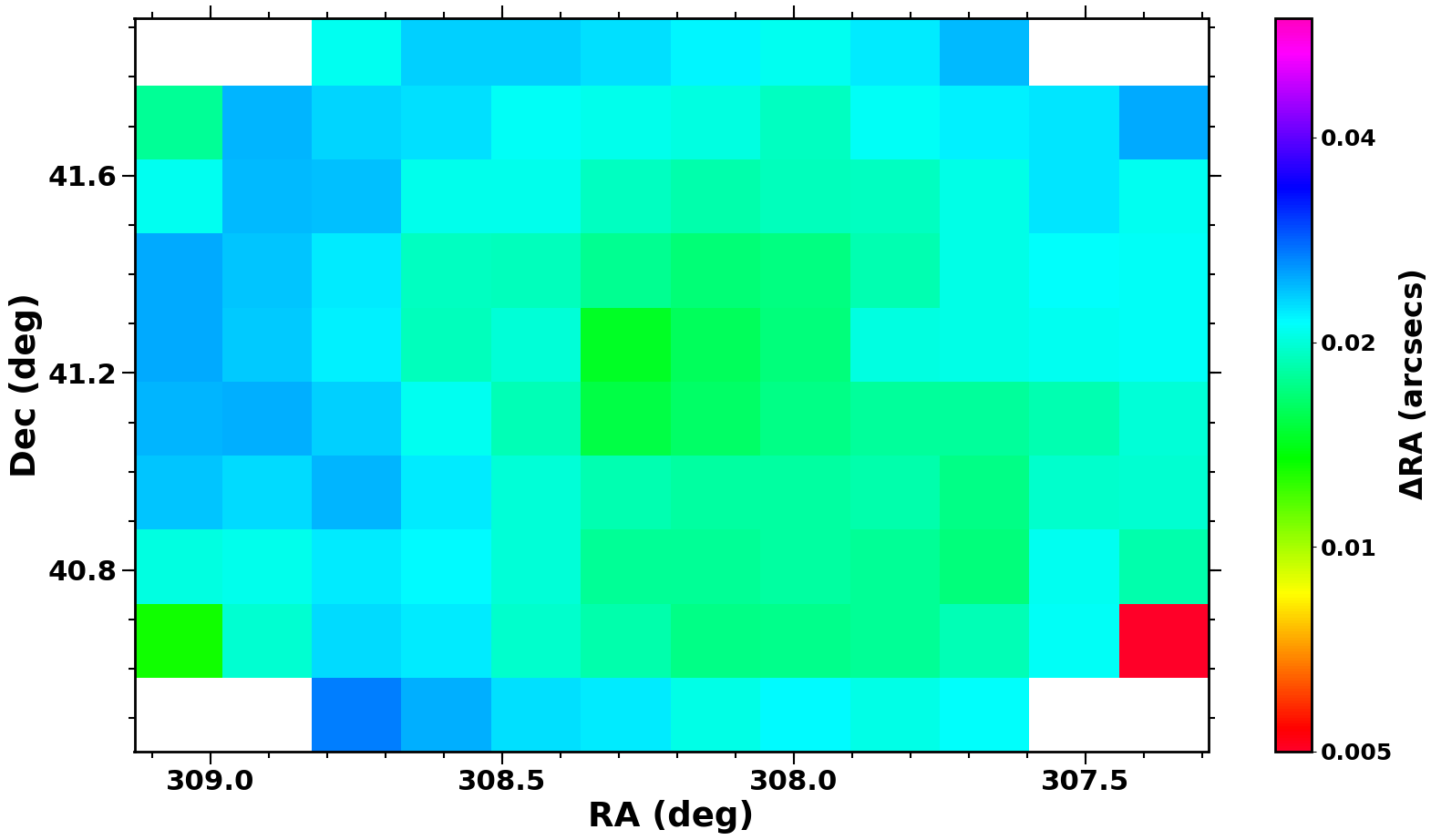}
	\includegraphics[width=8.6cm, height=6cm]{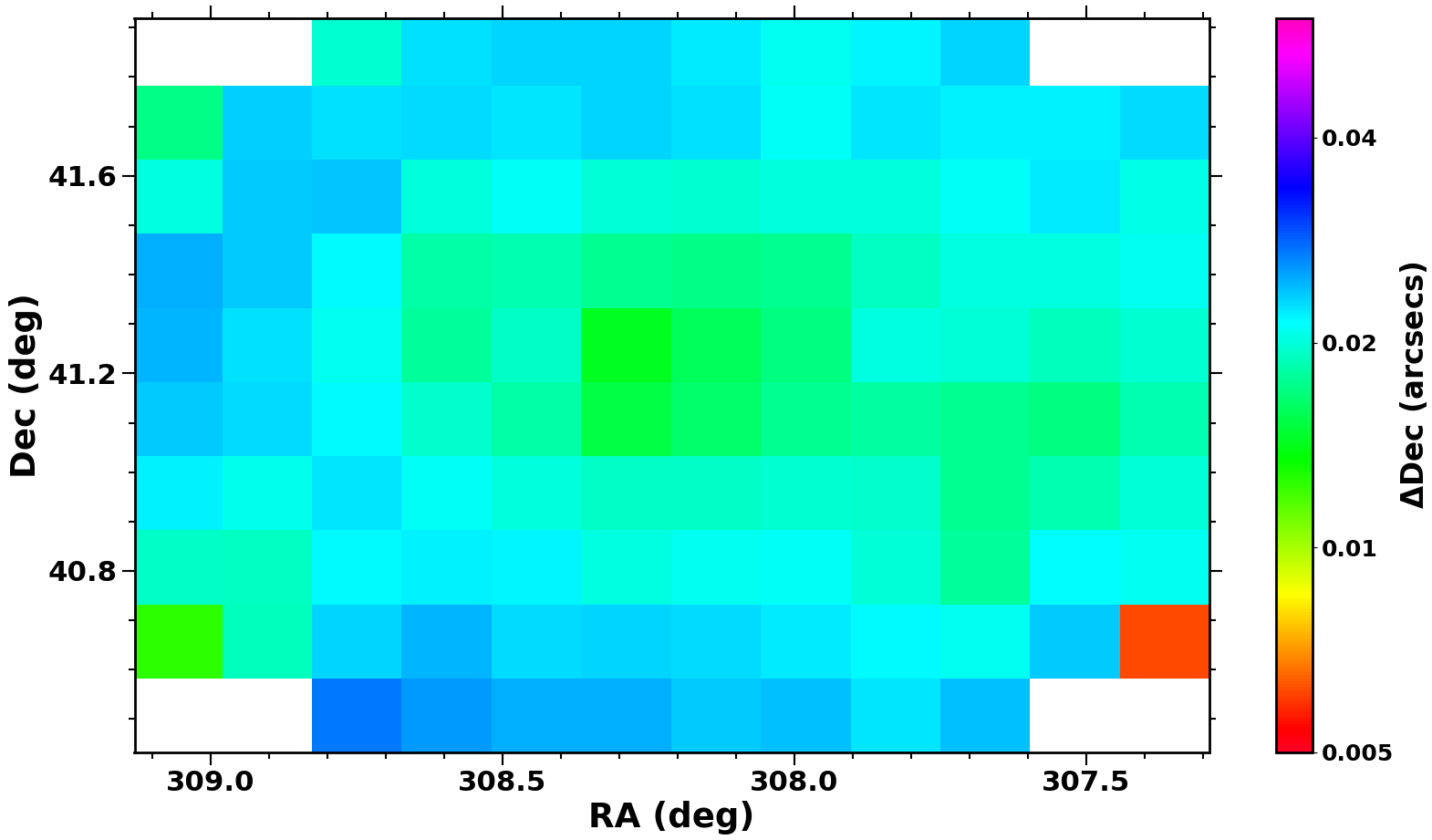}
	%\includegraphics[scale=0.18]{Del_RA_spatial.png}
	%\includegraphics[scale=0.18]{Del_Dec_spatial.png}
	%\includegraphics[scale = 0.2]{cyg_centre1.eps}
	%\includegraphics[scale = 0.38]{r2-Y_vs_r2.eps}%\scriptsize
	%\includegraphics[scale = 0.4]{z-K_vs_z.eps}%\scriptsize
	%\begin{small}  
	%\scriptsize
	%\linespread{0.8}
	\caption{ Spatial plots signifying the variation of internal error in Right Ascension ({\it Left}) and Declination ({\it Right}) across the entire region. The spatial maps are obtained by binning the RA and Dec parameter space into 10$^{\prime}$ $\times$ 10$^{\prime}$ bins across the entire observed region. The colorbar indicates the mean uncertainity in RA ({\it Left}) and Dec ({\it Right}) of each bin. The observed internal astrometric error ranges between 0.01$^{\prime\prime}$ - 0.03$^{\prime\prime}$, with almost uniform distribution throughout the region.
	 }%{\it Top left:}
% }
%\end{small}
\label{fig:astrometry spatial}
\end{figure*}

\begin{figure*}
	%\centering
%	\includegraphics[width=8.2cm, height=4.7cm]{Del_Ra_Hist.png}
%	\includegraphics[width=8.2cm, height=4.7cm]{Del_Dec_Hist.png}
	\includegraphics[scale=0.28]{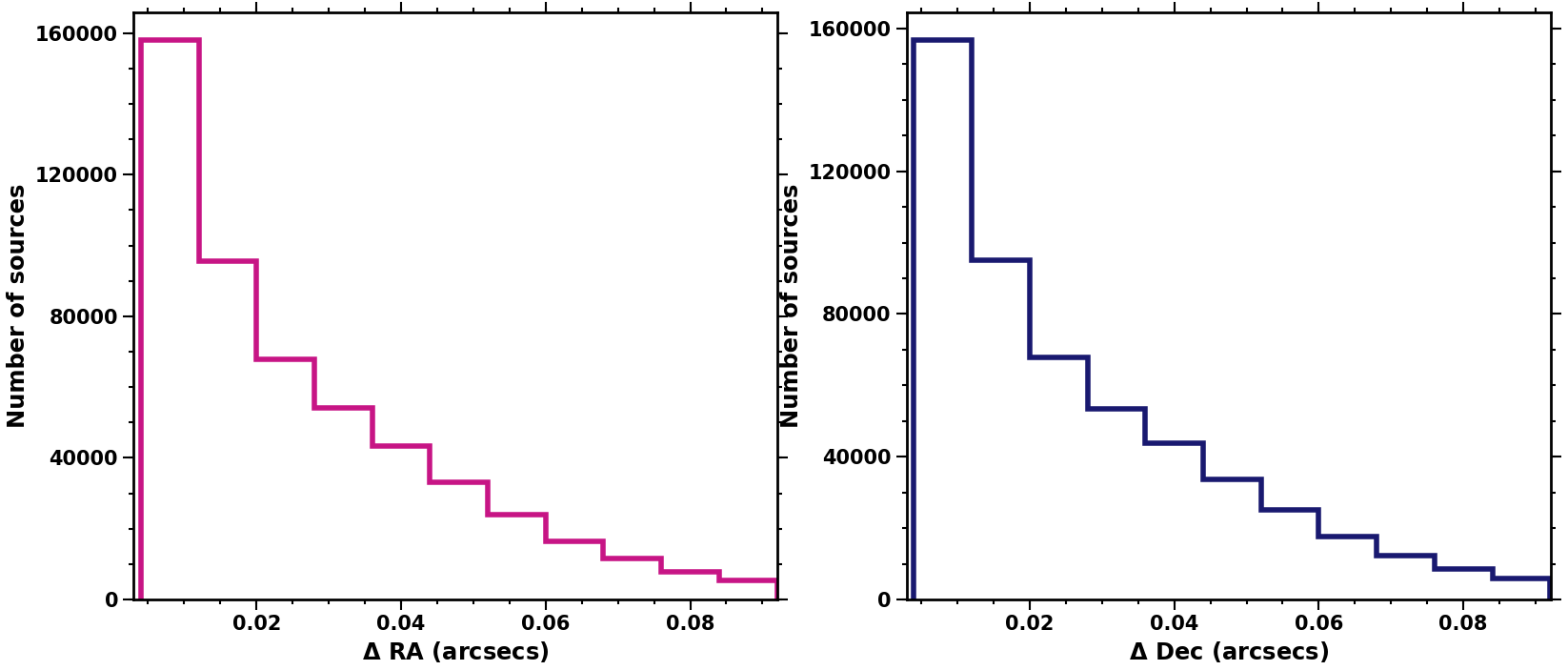}
%	\includegraphics[width=8.2cm, height=4.7cm]{Del_Dec_Hist.png}
	%\includegraphics[scale = 0.38]{r2-Y_vs_r2.eps}%\scriptsize
	%\includegraphics[scale = 0.4]{z-K_vs_z.eps}%\scriptsize
	%\begin{small}  
	%\scriptsize
	%\linespread{0.8}
	%\caption{ {\it Left} Histogram of internal error in Right Ascension and ({\it Right}) Declination. 
	% }
%{\it Top left:}
% }
%\end{small}
%\label{fig:astrometry hist}
\end{figure*}

\begin{figure*}
	%\centering
%	\includegraphics[width=8.2cm, height=4.7cm]{Del_Ra_Hist.png}
%	\includegraphics[width=8.2cm, height=4.7cm]{Del_Dec_Hist.png}
	\includegraphics[scale=0.25]{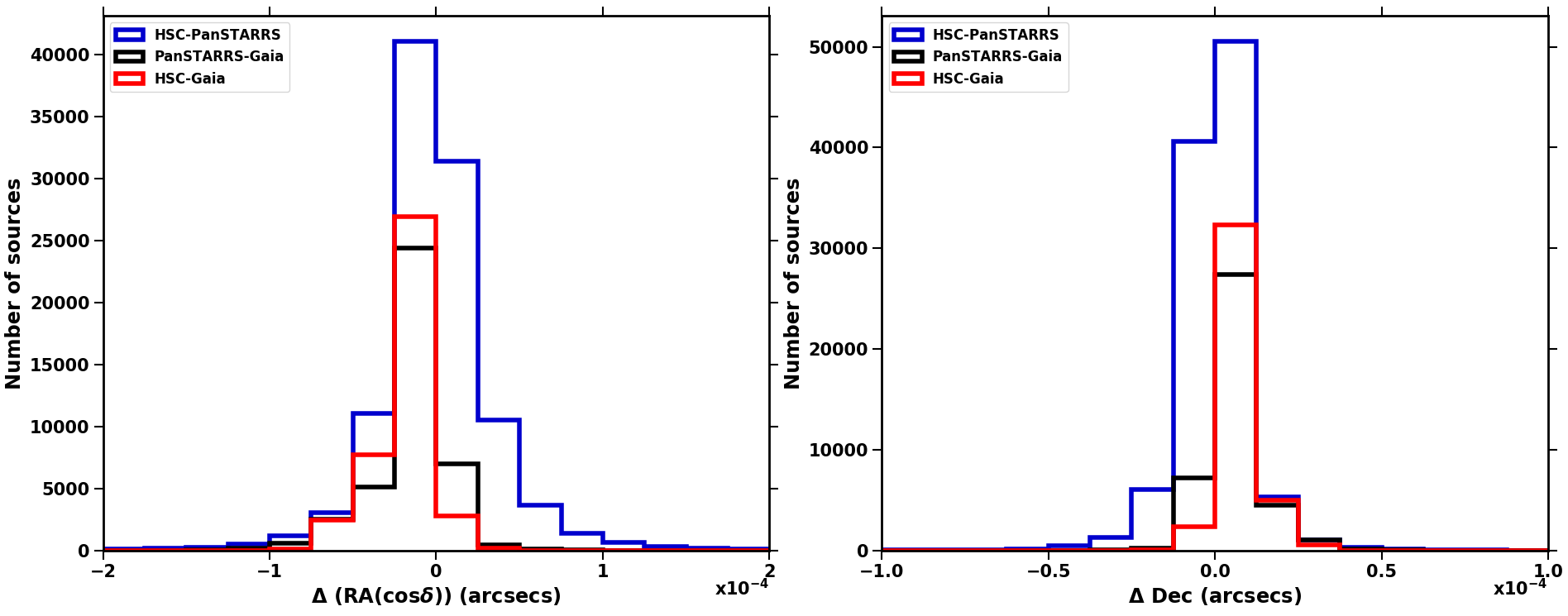}
%	\includegraphics[width=8.2cm, height=4.7cm]{Del_Dec_Hist.png}
	%\includegraphics[scale = 0.38]{r2-Y_vs_r2.eps}%\scriptsize
	%\includegraphics[scale = 0.4]{z-K_vs_z.eps}%\scriptsize
	%\begin{small}  
	%\scriptsize
	%\linespread{0.8}
	\caption{ {Histograms of internal error in Right Ascension (\it Top Left}) and  in Declination ({\it Top Right}). Histograms of the observed offset in  astrometry of HSC with respect to Pan-STARRS ({\it blue}), astrometry of Pan-STARRS with respect to Gaia EDR3 ({\it black}) and  astrometry of HSC with respect to Gaia EDR3 ({\it red}) in Right Ascension is shown in  the  {\it Bottom left} panel and in Declination is shown in the {\it Bottom Right} panel (See Section \ref{sec:astrometry} for details).
	 }
%{\it Top left:}
% }
%\end{small}
\label{fig:astrometry hist}
\end{figure*}

We present a graphical interpretation of the high precision astrometry of point sources in the HSC catalog in Figure \ref{fig:astrometry spatial} and \ref{fig:astrometry hist}. Due to our strict selection criteria (see Section \ref{sec:source select}), all the sources have both $\Delta$ RA and $\Delta$ Dec $\le$ 0.1$^{\prime\prime}$. This astrometric uncertainity of each source is attributed to the uncertainity in the position of its observed peak flux in different exposures. Hence, the mentioned astrometric error threshold of 0.1$^{\prime\prime}$ is a quality measure of the internal astrometric calibration relative to different visits. The internal astrometric error, mainly ranging between 0.01$^{\prime\prime}$--0.03$^{\prime\prime}$ appears to be uniform across the observed region (see Figure \ref{fig:astrometry spatial}) with a mean value $\sim$ 0.016$^{\prime\prime}$ for the detected sources. However, the census of sources decreases rapidly with increasing internal astrometric error (Figures \ref{fig:astrometry hist} {\it Top Left} and {\it Top Right}).\\

We perform an additional check of the astrometry of the detected HSC sources with respect to the external data such as Pan-STARRS DR1 and Gaia EDR3 available for the observed area of Cygnus OB2. The histograms in Figure \ref{fig:astrometry hist} {\it Bottom Left} and {\it Bottom Right} show the offset between HSC, Pan-STARRS DR1 and Gaia EDR3 astrometry in the HSC FOV (1.5$^\circ$ diameter region centred at Cygnus OB2). The absence of any visible offset between HSC and Pan-STARRS astrometry is attributed to the astrometric calibration performed with respect to Pan-STARRS PV3 DR1 data to develop a PSF model during the single-visit processing (refer Section \ref{sec: data reduction}). However, a  mean offset of $\sim 1.9 \pm 2^{\prime\prime} \times 10^{-5}$ in Right Ascension and $\sim 6.6 \pm 8^{\prime\prime} \times 10^{-6}$ in Declination is observed with respect to the Gaia EDR3 astrometry for both HSC and Pan-STARRS data and is well in accordance with the astrometric accuracy estimated in \citet{2019PASJ...71..114A} with these two data sets. We also present the spatial distribution of the astrometric offsets of HSC with respect to the GAIA EDR3 and Pan-STARRS DR1 astrometry in Figure \ref{fig:external astro spatial} and observe an excellent uniformity throughout the observed region.

\subsection{Completeness}
\label{sec:completeness}

\begin{figure}
	%\centering
	\includegraphics[width=\columnwidth, height=5.45cm]{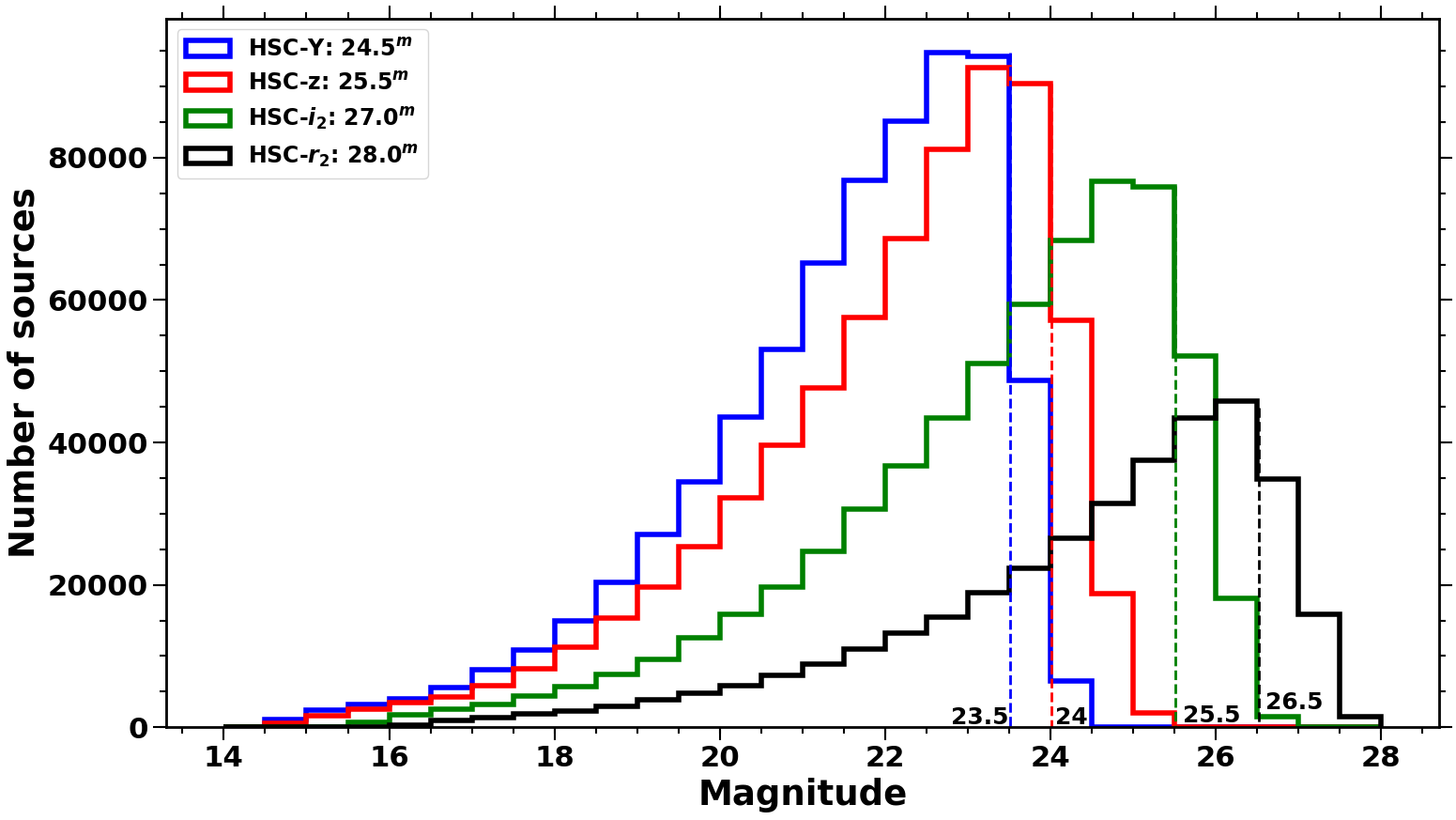}
	\includegraphics[width=1.05\columnwidth, height=5.5cm]{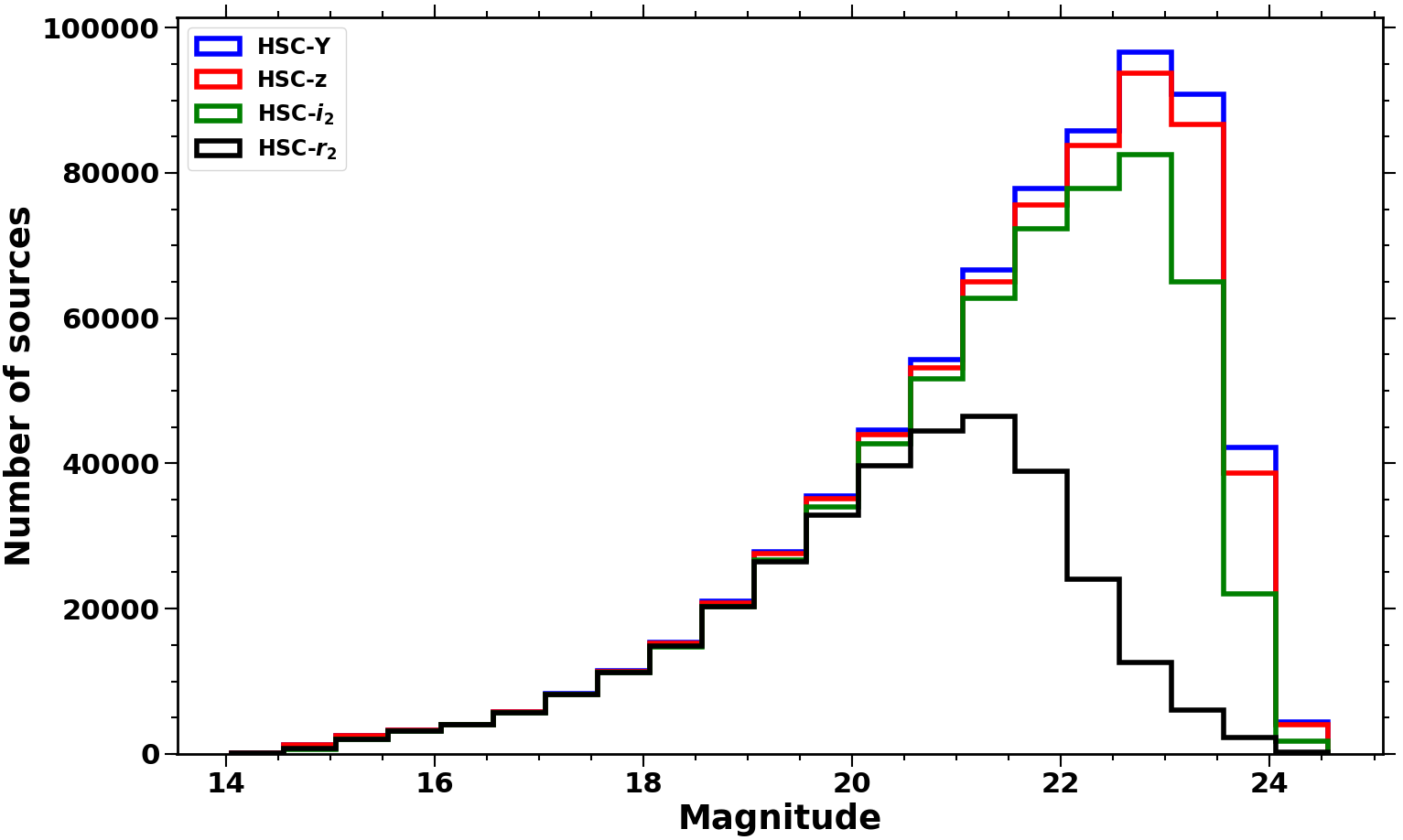}
%	\includegraphics[width=7.8cm, height=5cm]{Mag-lmt.png}
%	\includegraphics[width=8.2cm, height=4.6cm]{completeness.png}
	%\includegraphics[scale = 0.38]{r2-Y_vs_r2.eps}%\scriptsize
	%\includegraphics[scale = 0.4]{z-K_vs_z.eps}%\scriptsize
	%\begin{small}  
	%\scriptsize
	%\linespread{0.8}
	\caption{{\it Top}: Histograms representing the detection limit of individual HSC bands with {\it Black:} r$_{2}$-band; {\it Green:} i$_{2}$-band; {\it Red:} z-band and {\it Blue:} Y-band. The limiting magnitudes in individual HSC filters are mentioned in the legend. The dashed lines and the corresponding magnitudes denote the 90$\%$ completeness limit attained in individual filters as indicated by the turn-over point method (see Section \ref{sec:completeness} and Table \ref{tab:HSC analysis} for details). {\it Bottom}: Histogram depicting the completeness of r$_{2}$-band ({\it Black}) ; i$_{2}$-band ({\it Green}) and z-band ({\it Red}) with respect to the Y-band ({\it Blue}) of HSC.
	 } 
	 
%{\it Top left:}
% }
%\end{small}
\label{fig:completeness detect limit}
\end{figure}

\begin{figure}
	%\centering
%	\includegraphics[width=8.1cm, height=4.6cm]{compltns_compare.png}
	\includegraphics[width=0.953\columnwidth, height=5.45cm]{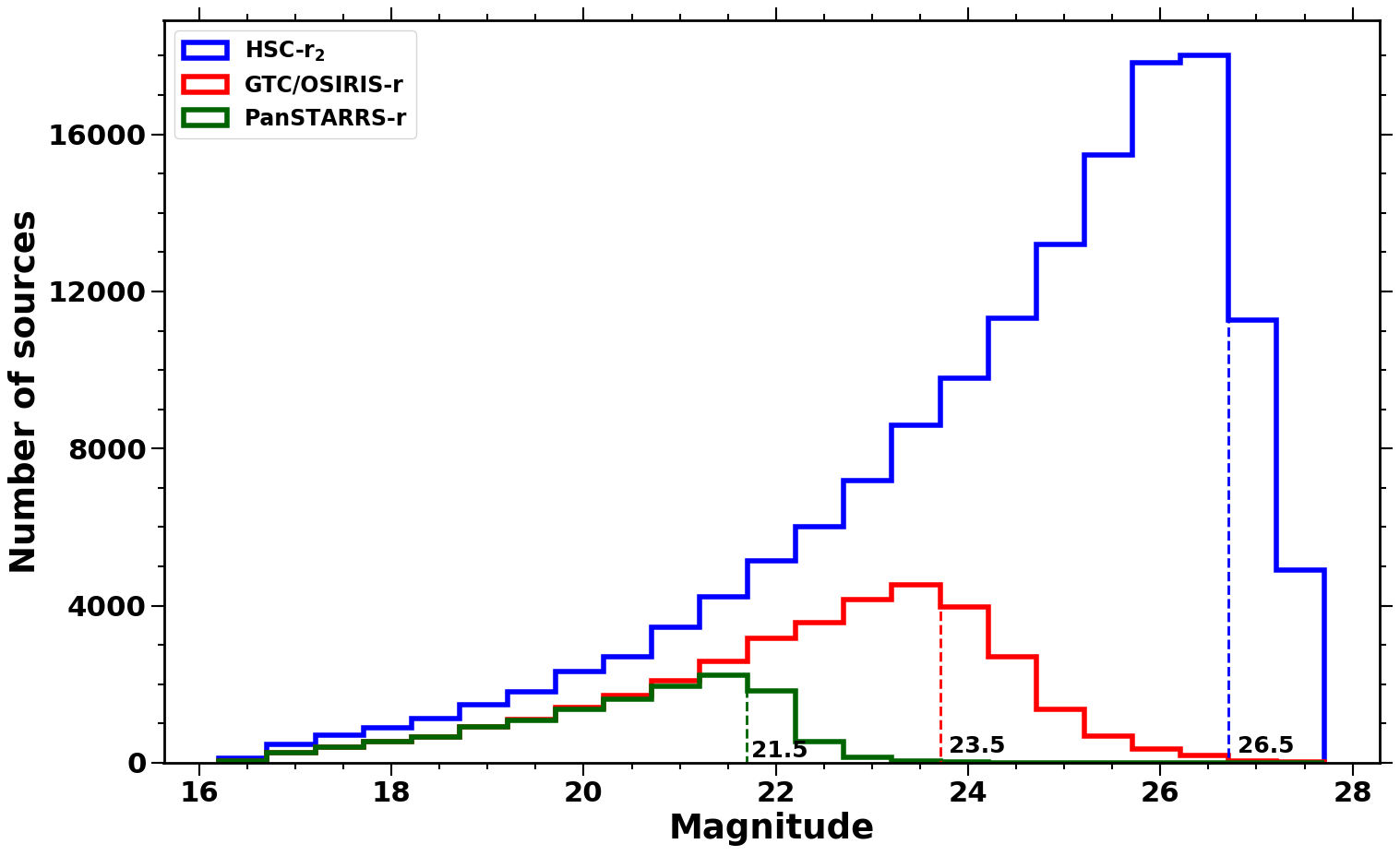}
	%\includegraphics[scale = 0.2]{cyg_centre1.eps}
	%\includegraphics[scale = 0.38]{r2-Y_vs_r2.eps}%\scriptsize
	%\includegraphics[scale = 0.4]{z-K_vs_z.eps}%\scriptsize
	%\begin{small}  
	%\scriptsize
	%\linespread{0.8}
	\caption{ Histogram plot representing the completeness of Pan-STARRS r-band ({\it Green}) and GTC/OSIRIS r-band ({\it Red}) with respect to the HSC $r_{2}$-band ({\it Blue}) for a comparable common area of 30$^{\prime}$ radius centred at Cygnus OB2. The dashed lines represent the corresponding 90$\%$ completeness limits which are found to be 21.5 mag for Pan-STARRS, 23.5 mag for GTC/OSIRIS and 26.5 mag for HSC.
	 }
%{\it Top left:}
% }
%\end{small}
\label{fig:completeness compare}
\end{figure}

The analysis of the final data gives the 5$\sigma$ limiting magnitude i.e the magnitude of the faintest star detectable with our observations in individual HSC filters. The histogram shown in Figure \ref{fig:completeness detect limit} ({\it Top}) indicates the detection limit of HSC photometry in different bands. In-spite of the high amount of nebulosity and moderate extinction prevalent in Cygnus OB2 (\citealt{2008MNRAS.386.1761D, 2010ApJ...713..871W, 2012ApJS..202...19G}), the limiting magnitude reaches down to\footnote{magnitude values rounded off to nearest 0.2 mag} 28.0 mag in r$_{2}$-band, 27.0 mag in i$_{2}$-band, 25.5 mag in z and 24.5 mag in Y-band. At a distance of 1600 pc, age $\sim$ 5 $\pm$ 2 Myrs (see Section \ref{sec: age}) and an average extinction A$_V$ ranging between 6 -- 8 mag (refer Section \ref{sec:CMD}), the mentioned detection limit of 27.0 mag in i$_{2}$-band corresponds to a stellar mass of 0.02 -- 0.03 M$_\odot$ (using isochrones of \citet{2015A&A...577A..42B}) i.e less than the Lithium-burning limit. The final HSC photometry is $\sim$ 90\% complete down to 26.5 mag, 25.5 mag, 24.0 mag and 23.5 mag in r$_{2}$, i$_{2}$, z and Y-band respectively, as indicated by the turn-over point in the histogram (denoted by dashed lines in Figure \ref{fig:completeness detect limit}). The turnover point in source count approach to evaluate the 90$\%$ completeness limit gives similar results to the artificial star-count method (\citealt{Jose_2016,2017ApJ...836...98J, 2021MNRAS.504.2557D,2021MNRAS.500.3123D}). Since Y-band has the highest number of detections, we take it as reference and calculate the number of counterpart sources in r$_{2}$, i$_{2}$ and z-band in each 0.5 mag bin to assess the completeness of other HSC filters relative to Y-band. The completeness of the photometry in various filters relative to Y-band attained by this method is presented in Figure \ref{fig:completeness detect limit} {\it Bottom}. We provide a summary of the useful quality parameters in individual HSC filters, for an age $\sim$ 5 $\pm$ 2 Myrs and A$_V$ = 6 - 8 mag in the Table \ref{tab:HSC analysis}. The obtained HSC photometry is found to be deeper by an order of 3 - 5 mag , when compared with the existing Pan-STARRS and GTC/OSIRIS photometry (limited to $\sim$ 21.5 mag and 23.5 mag, respectively in r-band), and thus provides a substantial sample of faint low mass sources in Cygnus OB2.\\

\begin{table*}
	\centering
	\begin{threeparttable}
	\caption{ Details of final HSC catalog in individual filters. (For more  details of the given parameters, please refer to Sections \ref{sec:photometric quality}, \ref{sec:astrometry}, \ref{sec:completeness} and \ref{sec:CMD})   }
	\label{tab:HSC analysis}
	\begin{tabular}{ |l|l|l|l|l| } % four columns, alignment for each
		\hline
		Filters & HSC-Y & HSC-z & HSC-$i_{2}$ & HSC-$r_{2}$ \\
		\hline
		Number of sources & 699,798 & 685,511 & 622,011 & 358,372 \\
		Fraction of sources $\le$ 0.05 mag error & 91\% & 95\% & 93\% & 90\% \\
		%Magnitude Offset & 0.03$\pm$0.06 & 0.01$\pm$0.07 & 0.01$\pm$0.02 & 0.01$\pm$0.03 \\
		Brightness limit\tnote{a} (mag) & 14.0 & 14.2 & 15.3  & 15.6  \\
		Limiting magnitude\tnote{b,c} (mag) & 24.5  & 25.5  & 27.0  & 28.0  \\
		Limiting magnitude upto 90\% completeness (mag) & 23.5  & 24.0  & 25.5  & 26.5  \\
		Limiting mass (in M$_\odot$)\tnote{d} & 0.02-0.03 & 0.03-0.04 & 0.03-0.06  & 0.15-0.30 \\
		\hline
	%\caption{Table 1 shows various parameters obtained by analysis of final HSC catalog in individual filters. Here}
	%\label{tab:HSC Analysis}
	\end{tabular}
	\begin{tablenotes}\footnotesize
	\item [a] Magnitude of the brightest object detected
	\item [b] Magnitude of the faintest object detected
	\item [c] Magnitudes rounded off to 0.2 mag
	\item [d] Mass corresponding to magnitude with 90\% completeness for A$_V$: 6 -- 8 mag and age: 5 $\pm$ 2 Myrs.
	\end{tablenotes}
	\end{threeparttable}
\end{table*}

\section{Data Analysis and Results}
\label{sec:analysis}

We present here some preliminary analysis based on the HSC data to illustrate the significance of Cygnus OB2 as an ideal target for low-mass star formation studies with the help of a few color-magnitude diagrams (CMDs) presented in this section. We also perform a statistical field decontamination using a peripheral control field to obtain a statistical estimate of member stars and use that to obtain the approximate median age and average disk fraction of the central 18$^\prime$ region of Cygnus OB2. \\

\subsection{Color-Magnitude Diagrams}
\label{sec:CMD}

\begin{figure}
	%\centering
	%\includegraphics[width=8.5cm, height=6cm]{r2z_cmd_whole.png}
%	\includegraphics[width=8.5cm, height=6cm]{r2z_cmd_whole.png}
	%\includegraphics[scale=0.19]{zy_cmd_whole.png}
	\includegraphics[scale=0.26]{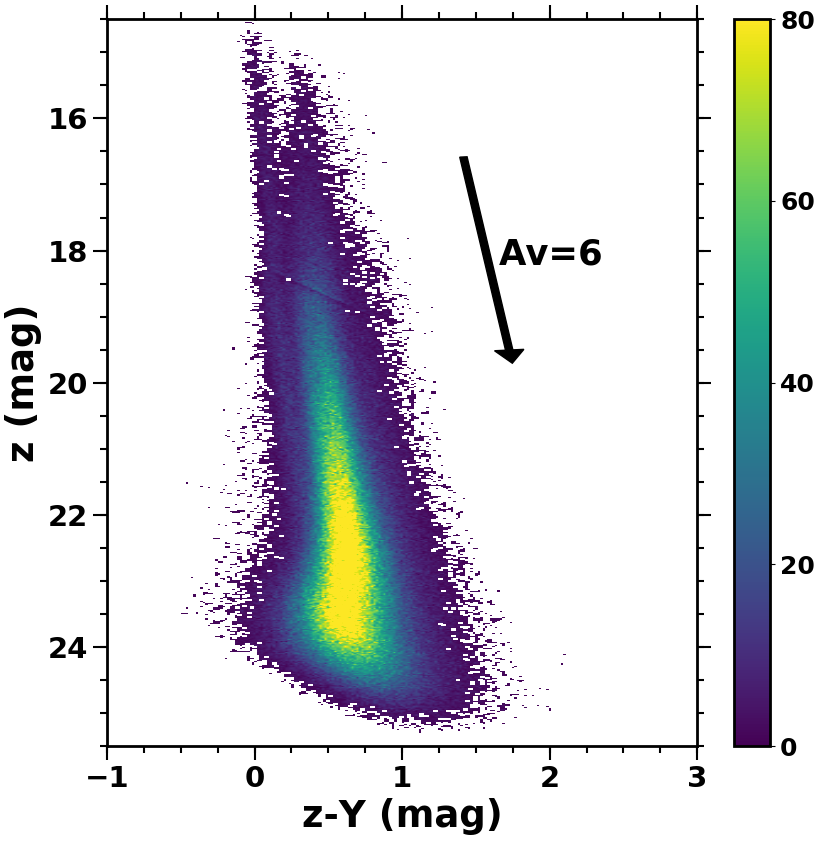}
	%\includegraphics[scale = 0.38]{r2-Y_vs_r2.eps}%\scriptsize
	%\includegraphics[scale = 0.4]{z-K_vs_z.eps}%\scriptsize
	%\begin{small}  
	%\scriptsize
	%\linespread{0.8}
	\caption{ Hess plot of z-Y vs  z Color-Magnitude Diagram (CMD) with HSC sources detected in the entire area of 1.5$^\circ$ diameter centred at Cygnus OB2. The Hess plot is obtained by binning the color and magnitude parameter space into bins of size 0.01 mag and 0.03 mag respectively. The black arrow marks the direction of reddening vector of $A_V$ = 6 mag.
	 } 
	 
%{\it Top left:}
% }
%\end{small}
\label{fig:whole cmd}
\end{figure}

\begin{figure*}
	%\centering
	%\includegraphics[width=8.5cm, height=6cm]{i2Y_cmd_Yso.png}
	\includegraphics[scale=0.19]{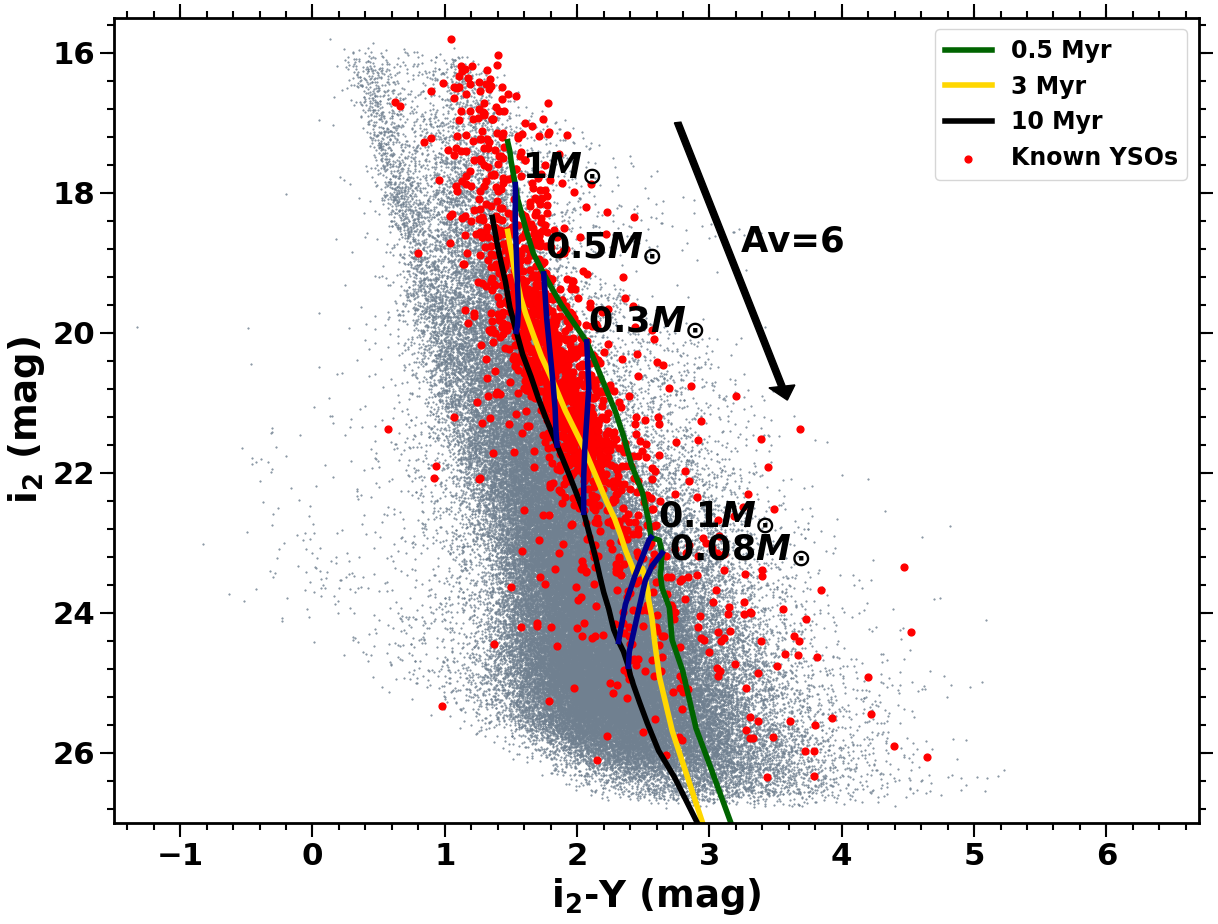}
	\includegraphics[scale=0.19]{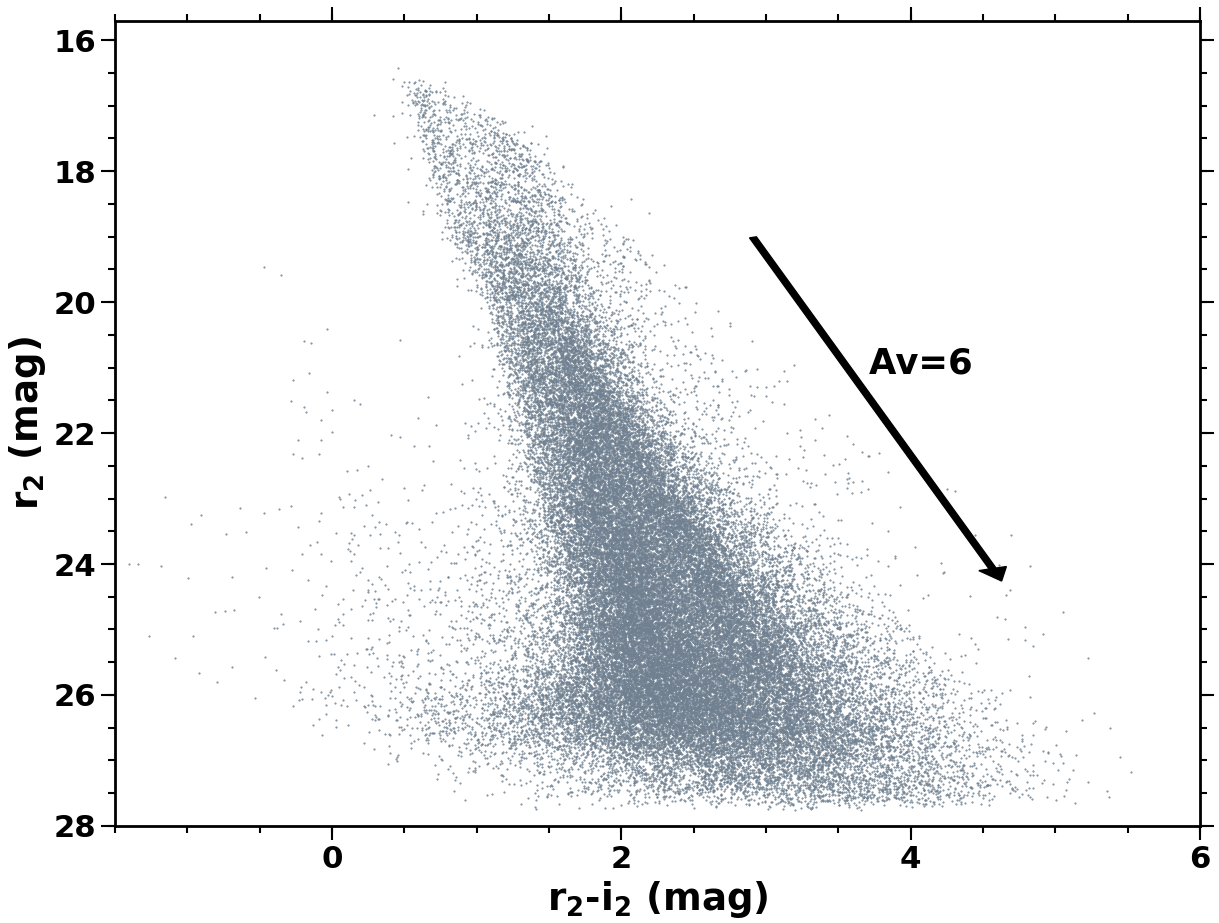}
	%\includegraphics[scale = 0.38]{r2-Y_vs_r2.eps}%\scriptsize
	%\includegraphics[scale = 0.4]{z-K_vs_z.eps}%\scriptsize
	%\begin{small}  
	%\scriptsize
	%\linespread{0.8}
	\caption{ {\it Left}: i$_{2}$-Y vs i$_{2}$ CMD within the central 18$^\prime$ radius of Cygnus OB2.  Isochrones of age 0.5, 3 and 10 Myr  and evolutionary tracks for various masses \citep{2015A&A...577A..42B}, which are corrected for an Av=6 mag and distance = 1600 pc are shown using solid curves.  The previously known YSOs of the complex  (\citealt{Guarcello_2013}) are overplotted as red dots. {\it Right}: r$_{2}$-i$_{2}$ vs r$_{2}$ CMD for the same 18$^\prime$ radius region. The black arrow marks the direction of reddening vector for A$_V$ = 6 mag.
	 } 

%{\it Top left:}
% }
%\end{small}
\label{fig:CMD}
\end{figure*}

Color-magnitude diagrams (CMDs) are integral to segregate the cluster members from foreground and background contaminants (e.g \citealt{2017ApJ...836...98J, Esplin_2020, 2021MNRAS.504.2557D}) and estimate the age, temperature and spectral type of member stars in a star-forming cluster. We present the Hess plot of the z-Y vs z Color-Magnitude Diagram (CMD) in Figure \ref{fig:whole cmd}, plotted with our optical catalog obtained for the entire 1.5$^\circ$ diameter area of Cygnus OB2. A similar i$_{2}$-Y vs i$_{2}$ CMD in Figure \ref{fig:CMD} ({\it Left}) and r$_{2}$-i$_{2}$ vs r$_{2}$ CMD in Figure \ref{fig:CMD} ({\it Right}) have been plotted for the sources lying in the central region of 18$^\prime$ radius. This area has been particularly selected due to the high concentration ($\sim$ 50$\%$ of the total) of YSOs (identified previously by \citet{Guarcello_2013}) present in this region. Cygnus OB2 exhibits a distinct pre-main sequence branch which is a prominent feature observed in CMDs of young clusters (\citealt{2013MNRAS.432.3445J,2017ApJ...836...98J, Panwar_2018, 2019A&A...623A.112D, 2019ApJ...875...51B, 2020arXiv201200524K, 2021MNRAS.504.2557D}). In order to analyse the approximate age of the cluster, we over-plot isochrones of age 0.5, 3 and 10 Myr and evolutionary tracks for various masses from \citet{2015A&A...577A..42B} on the i$_{2}$-Y vs i$_{2}$ CMD. As per the past studies, an extinction of A$_V$ = 4 -- 5 mag has been observed towards the north-west of Cygnus OB2 along with A$_V$ = 5.5 -- 7.0 mag observed towards centre and south of the association (\citealt{10.1093/mnras/stv323}). Hence, we choose a mean value of extinction as A$_V$ = 6.0 mag in order to redden our isochrones. The isochrones have been reddened using the extinction laws of \citet{Wang_2019} for Pan-STARRS filter system, taking A$_V$ = 6.0 mag and 1600 parsecs as distance of Cygnus OB2 from the Sun (\citealt{10.1093/mnras/stz2548}). % the average of 1350 parsecs and 1755 parsecs i.e distances to the two substructures in Cygnus OB2 as given by \citealt{2019MNRAS.484.1838B}, is $\sim$ 1600 parsecs). 
Consequently, the transformation equations (given in Appendix \ref{sec:transform eq}) have been used to convert the obtained magnitudes of Baraffe isochrones (in Pan-STARRS filter system) to HSC filter system.\\
 
The majority ($\sim 88\%$) of the previously detected YSOs (\citealt{Guarcello_2013}), overplotted as red circles, are located within the 10 Myr isochrone overplotted on the i$_{2}$-Y vs i$_{2}$ CMD in Figure \ref{fig:CMD} {\it{Left}} and thus, occupy the characteristic pre-main sequence branch. The source population occupying the young pre-main sequence branch consists of both cluster members as well as background contaminants. We obtain a statistical estimate of the membership in the central 18$^\prime$ using the field decontamination process further in Section \ref{sec:field decontamination}. The color of these sources (i.e i$_{2}-Y \ge$ 2) reinforces the claim that they constitute the pre-main sequence population present in the central 18$^\prime$ radius region of Cygnus OB2.  % To accentuate this feature, we have additionally presented in Figure \ref{fig:CMD} {\it Bottom Right} the corresponding hess plot of the $i_{2}$-Y vs $i_{2}$ CMD plotted for the same 20' radius region. The hess plot is generated by dividing the color-magnitude plot into square bins of size 0.05 mag and estimate the number of sources falling in each bin. The color bar suggests the density of sources in each bin of the color-magnitude diagram. The distinct branching of pre-main sequence sources $<$ 10 Myr appears vividly in the hess plots. 
 %We find that approximately, 45$\%$ of the total sources in the entire observed 1.5$\circ$ diameter region of Cygnus OB2 have an age $\le 10 Myrs$ and a significant fraction of this young population lies in the sub-stellar regime ($\le 0.07 M_\odot$), as inferred by the isochrone fitting. However, the mentioned fraction (45$\%$) $\le 10 Myrs$ is an upper limit on the pre-main sequence population in the region as this sample may contain many foreground and background contaminants.\\
 
\begin{figure*}
	%\centering
%	\includegraphics[width=8.5cm, height=6cm]{r2y_inner_cmd.png}
	\includegraphics[scale=0.275]{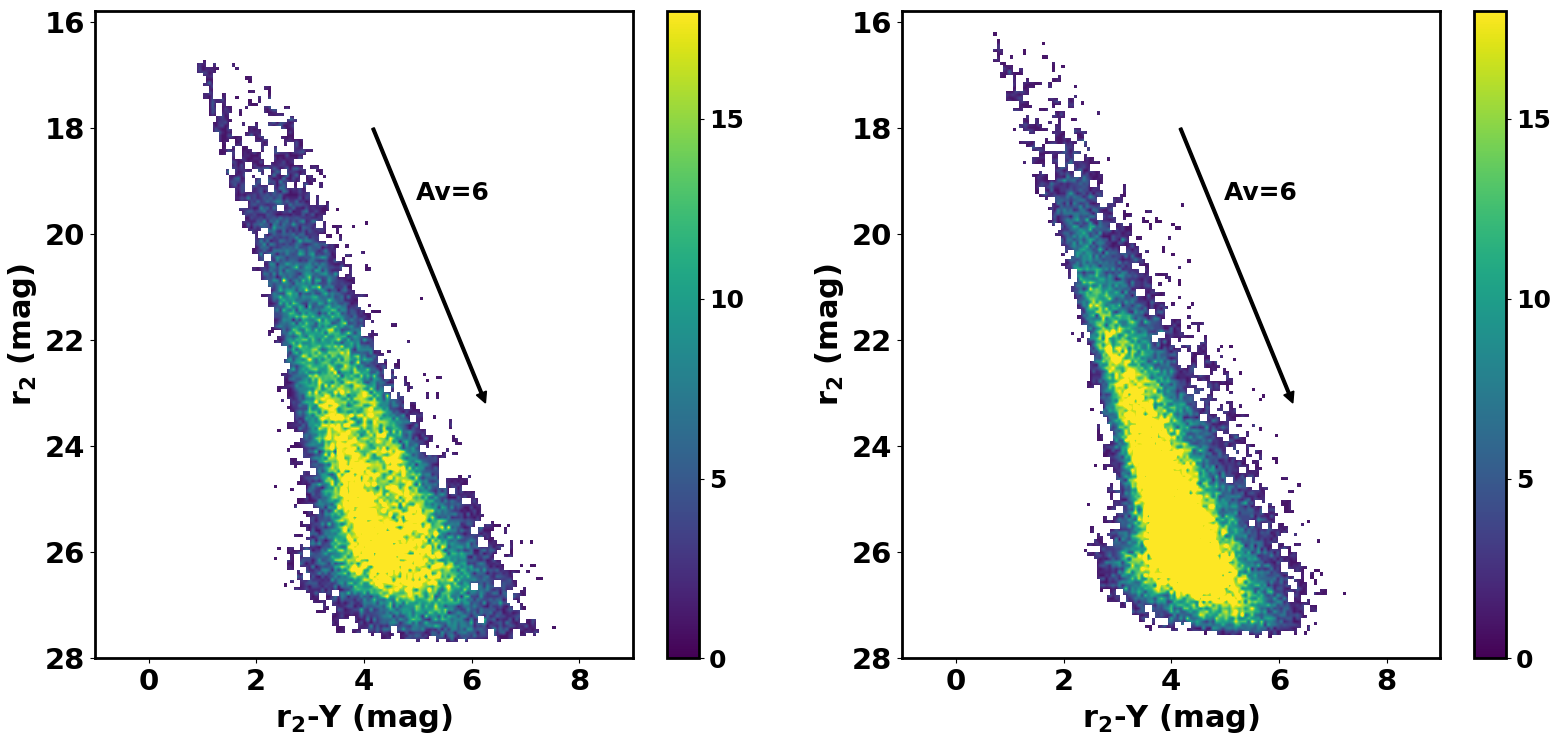}
	%\includegraphics[scale=0.198]{r2y_inner_cmd_hess.png}
	%\includegraphics[scale=0.198]{r2y_outer_cmd_hess.png}
	%\includegraphics[scale = 0.38]{r2-Y_vs_r2.eps}%\scriptsize
	%\includegraphics[scale = 0.4]{z-K_vs_z.eps}%\scriptsize
	%\begin{small}  
	%\scriptsize
	%\linespread{0.8}
	\caption{ The comparative Hess diagrams of r$_{2}$-Y vs r$_{2}$ CMDs to emphasize cluster membership for sources located within ({\it Left}) the inner 18$^\prime$ radius of Cygnus OB2 (RA: 308.2785; Dec: 41.7477) and {(\it Right}) a rectangular region of the same area towards the outskirts of Cygnus OB2 (RA: 308.2655; Dec: 41.7497). The black arrow marks the direction of reddening vector for A$_V$ = 6 mag.
	 }
	 
%{\it Top left:}
% }
%\end{small}
\label{fig:CMD membership}
\end{figure*}
We emphasize the cluster membership of the sources in the pre-main sequence branch with the aid of a comparative study between an 18$^\prime$ radius circular region towards the centre and an equal rectangular area towards the periphery of Cygnus OB2 (RA: 308.2665; Dec: 41.7497), as shown in Figure \ref{fig:CMD membership}. The Hess plot of r$_{2}$-Y vs r$_{2}$ CMD (Figure \ref{fig:CMD membership} ({\it Left})) is plotted for the sources in the central 18$^\prime$ radius region, which is prolific in pre-main sequence cluster members and a similar Hess plot is plotted in Figure \ref{fig:CMD membership} ({\it Right}) for the sources lying towards the outskirts of Cygnus OB2. The absence of a distinguished pre-main sequence branch in the CMD of the sources towards the periphery as compared to the central region, suggests that it is mainly populated by the non-cluster members in the  foreground or background. Hence, in accord with the literature (\citealt{2000A&A...360..539K, 2010ApJ...713..871W, Guarcello_2013, 10.1093/mnras/stv323, 2016arXiv160501773G}), our optical data analysis advocates that Cygnus OB2 is an active young star formation site rich in pre-main sequence, low mass as well as sub-stellar population with a suggested age $\le$ 10 Myrs. \\

\subsection{Field Star Decontamination}
\label{sec:field decontamination}
The background and foreground contaminants, also termed as field star contaminants, generally lie in the line of sight of the observed target region and can overlap with the young pre-main sequence population in the CMDs as mentioned in Section \ref{sec:CMD}. Hence, the identification of cluster members is particularly crucial for an accurate estimation of various cluster parameters like age, distance, disk fraction which can otherwise be biased by the presence of field stars. Although, kinematic parameters like proper motion, radial velocity and other methods such as spectroscopy and SED analysis provide the most precise membership identification (\citealt{2017MNRAS.468.2684P, 2018MNRAS.476.2813D, 2019ApJ...878..111H, 2019A&A...627A.135B, 2020ApJ...892..122J, 2021MNRAS.500.3123D}), such data is available only for a handful of the sources with Gaia eDR3 counterparts complete down to $\sim$ 20 mag, which is inadequate for the low mass pre-main sequence members in Cyngus OB2. Hence, a statistical field star subtraction using an appropriate control field is useful to obtain a statistical estimate of the probable cluster members down to faint low mass limits (r$_{2}$ $\sim$ 28 mag) (eg. \citealt{2017ApJ...836...98J, 2020ApJ...896...29K, 2021MNRAS.504.2557D}). \\

\begin{figure*}
	%\centering
%	\includegraphics[width=8.5cm, height=6cm]{r2y_inner_cmd.png}
%	\includegraphics[width=8.5cm, height=6cm]{r2y_inner_cmd_mod.png}
	\includegraphics[scale=0.33]{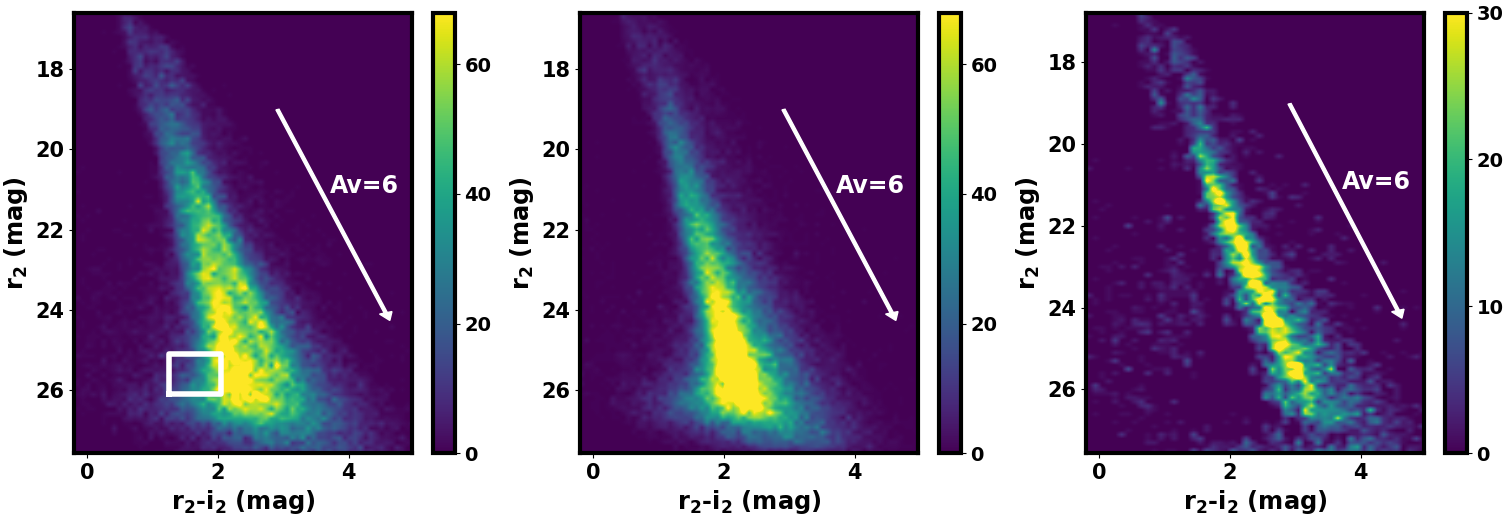}
	%\includegraphics[scale=0.19]{r2y_outer_cmd_new.png}
	%\includegraphics[scale = 0.38]{r2-Y_vs_r2.eps}%\scriptsize
	%\includegraphics[scale = 0.4]{z-K_vs_z.eps}%\scriptsize
	%\begin{small}  
	%\scriptsize
	%\linespread{0.8}
	\caption{ Hess plots of r$_{2}$-i$_{2}$ versus r$_{2}$ CMD for {\it (Left)} the cluster field, {\it (Middle)} the control field and {\it (Right)} the field subtracted sources. For the hess plot of  control field (Middle), the control field data count per bin is scaled by the median log likelihood value, i.e 0.73. A sample box area chosen to calculate this log likelihood value is shown as the white box in the Hess plot of the cluster field {\it (Left)}. Several such box areas are considered to calculate the median log likelihood value. The white arrow marks the direction of reddening vector for A$_V$ = 6 mag.
	 }
	 
%{\it Top left:}
% }
%\end{small}
\label{fig:field sub}
\end{figure*}

We perform the statistical field decontamination for a cluster field of 18$^\prime$ radius centred at Cygnus OB2, which encloses $\sim$ 50$\%$ of the known YSOs in the region. In the absence of a control field external to the observed region, we choose a rectangular control field located towards the outskirts of the Cygnus OB2 (centred at RA: 308.2655; Dec: 41.7497) of an area equal to that of the cluster field. This control field is the same as used above for Figure \ref{fig:CMD membership} {\it{(Right)}}. We observe a higher source density in the control field as compared to the cluster field, which may either be due to differences in the stellar density or could be attributed to the different extinction observed in the two directions. Although, the CO maps and mid-IR images from MSX from \citet{2006A&A...458..855S} and \citet{2008hsf1.book...36R} suggest an approximate uniform extinction across the Cygnus OB2, the extinction mapping performed by us using deep near-IR UKIDSS data (to be discussed in the forth-coming work.) reveals moderate differential reddening across the region with the control field being less extincted than the cluster field by 1 - 1.5 mag. To address the stellar density fluctuation, we chose a box in the color magnitude diagram where we do not expect to see any pre-main sequence stars in the cluster field (such as the one shown in Figure \ref{fig:field sub} {\it (Left)}). We scale down the counts in the color magnitude diagram of the control field by a constant factor $f$, such that the number of detected objects in this box is consistent between the cluster and the control field within Poisson fluctuations. We infer the posterior distribution of the parameter $f$ using Monte Carlo Markov sampling using the package \textsc{emcee} \citep{2013PASP..125..306F}. 
%difference in the observed source density and hence, t  avoid the over subtraction of the sources, we apt for log likelihood algorithm\footnote{emcee python package\citealt{2013PASP..125..306F}} with Poisson distribution. 
We performed multiple iterations over several smaller box areas (located over the entire r$_{2}$ magnitude range and r$_{2}$ - i$_{2}$ color $\le$ 2) in the CMD of the control field, and obtain a median likelihood value of 0.73 that is used to scale the bin counts of the control field in the entire color magnitude diagram. This median likelihood value scales down the overdensity of sources in the control field, which can otherwise result in the over subtraction of the sources while performing field decontamination of the cluster field.\\%The extinction across the cluster and the control field can be considered almost uniform as evident by the CO maps and mid-IR images from MSX along with the extinction mapping with near-IR 2MASS data from \citet{2006A&A...458..855S} and \citet{2008hsf1.book...36R}. Both cluster and the control field regions have similar dust and gas distributions, hence do not suffer significantly from the problem of differential reddening.}}\\

We then perform the field subtraction using r$_{2}$-i$_{2}$ versus r$_{2}$ CMD and divide the color and magnitude parameter space into 0.1 and 0.2 mag bins. For each bin, we first scale down the count of sources in the control field and then, subtract the control field count from the cluster field count. The resultant count thus obtained, is a floating point number which represents the average number of sources to be selected randomly as the field subtracted sources in each bin. Hence, in order to obtain an integer count, we randomly select an integer value within the Poisson fluctuations of the average count obtained as a result of subtraction. The derived integer count is considered as the number of sources to be selected as field subtracted sources in the cluster field per bin. %The field subtracted sources, equal to the derived integer count in number, are selected randomly in the cluster field per bin. is considered as the number of sources to be selected as field subtracted sources chosen randomly in the cluster field per bin. We then, randomly select these field subtracted sources in the cluster field. The resultant count, thus obtained after scaling by factor f is rounded off to the nearest integer. We then, perform one to one matching between the sources in control and cluster field for each bin, based on their color and magnitudes. For each source in the control field, its nearest match found on the basis of its r$_{2}$-i$_{2}$ color and r$_{2}$ magnitude, is removed from the cluster field. The one to one matching simply selects the sources which should be removed from each bin in cluster field }} This process is repeated for all the bins, which leaves us with field subtracted sources in the cluster field.
We emphasize here that this field decontamination is purely statistical and the resultant field subtracted sources may not be the confirmed members of the cluster. The Figure \ref{fig:field sub} shows the Hess plots of r$_{2}$-i$_{2}$ versus r$_{2}$ CMD for the cluster and control field along with that for the field subtracted sources. We observe that the field subtracted sources distinctly occupy the pre-main sequence branch in the CMD with a few scattered sources, which can be attributed to the statistical uncertainty in the field decontamination process. We repeated the field subtraction with another control field located in the outskirts of Cygnus OB2, and find that the statistics remain comparable within 10$\%$ uncertainty. Hence, we consider the field subtracted sources for further analysis to estimate the median age and disk fraction of the chosen cluster field area as described in the following sections.\\ 

\subsection{Age distribution of Cygnus OB2}
\label{sec: age}
\begin{figure*}
	%\centering
	%\includegraphics[width=8.5cm, height=6cm]{i2Y_cmd_Yso.png}
	\includegraphics[scale=0.25]{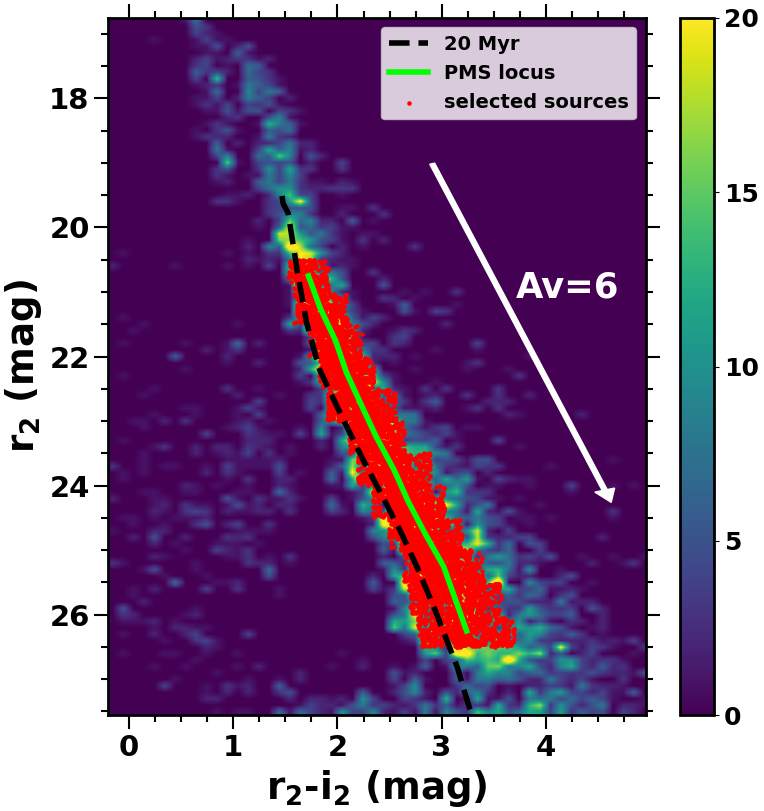}
	\includegraphics[scale=0.2]{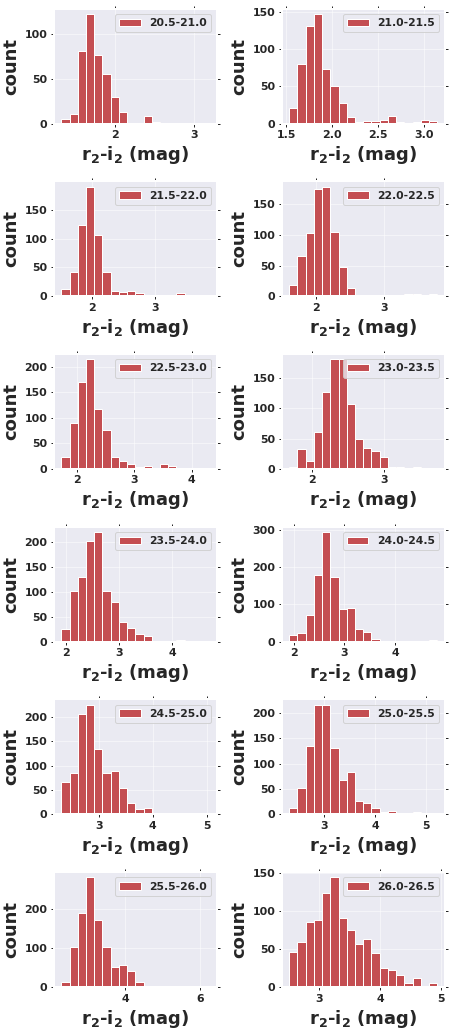}
	%\includegraphics[scale = 0.38]{r2-Y_vs_r2.eps}%\scriptsize
	%\includegraphics[scale = 0.4]{z-K_vs_z.eps}%\scriptsize
	%\begin{small}  
	%\scriptsize
	%\linespread{0.8}
	\caption{{\it Left}: Hess plot of r$_{2}$ - i$_{2}$ vs r$_{2}$ CMD of the field subtracted members in the central cluster field of 18$^\prime$ region of Cygnus OB2.  This is overplotted with the selected sources (red dots) i.e within 1 $\sigma$ limits of the empirical pre-main sequence (PMS) locus (green solid curve) and 20.5 mag $\le$ r$_{2}$ $\le$ 26.5 mag. These selected sources are considered for the age estimation. Also, the 20 Myr Baraffe isochrone corrected for an Av=6 mag and distance = 1600 pc is shown as the black dashed curve. The white arrow marks the direction of reddening vector for A$_V$ = 6 mag. {\it Right}: Histograms for r$_{2}$ - i$_{2}$ color distribution in each r$_{2}$ magnitude bin of 0.5 mag (the legend in each histogram shows the respective magnitude bin for which the histogram of color distribution is plotted).% The selected sources in each bin i.e within 1 $\sigma$ limits of median r$_{2}$ - i$_{2}$ color, are fitted with the Gaussian, the peak of which defines the mean PMS locus.}} 
	 } 

%{\it Top left:}
% }
%\end{small}
\label{fig:age bin}
\end{figure*}

The information about the age of the sources, combined with an estimate of the disk bearing sources (YSOs) in a cluster is helpful in constraining the star formation history of the region. However, the age estimation can be biased if the sample is contaminated with field stars. Hence, we use the statistically subtracted sources obtained after the field decontamination process, described above in Section \ref{sec:field decontamination}, to estimate the age of the chosen cluster field area. However, to eliminate any leftover contaminants due to statistical error in the field decontamination process which may bias our age estimation, we consider only those sources with 20.5 mag $\le$ r$_{2}$ $\le$ 26.5 mag, in accordance with the completeness limit of r$_{2}$-band. The upper limit of 20.5 mag corresponds to 1.4 M$_\odot$ source (the upper mass limit in Baraffe isochrones) at an age $\sim$ 5 Myrs. Since, approximately 90$\%$ of the total field subtracted sources have mass less than the considered upper limit, it will not modify our results significantly. %We consider only those decontaminated sources which are located leftwards of the 0.5 Myr isochrone, the minimum age for which the Baraffe isochrone models are available and lose mere $\sim$ 2$\%$ of the total sources with this criteria in the defined magnitude range. 
To further refine our selection, we define an empirical pre-main sequence (PMS) locus and select only those sources which are within 1 $\sigma$ limits of this empirical locus. We refer to these sources as the selected sources. The PMS locus is obtained by dividing the r$_{2}$ magnitude range into 0.5 mag bins. For each bin then, we take the mean of the r$_{2}$ magnitude and median of the r$_{2}$ - i$_{2}$ color of the sources inside the bin. This mean magnitude and the median r$_{2}$ - i$_{2}$ color in each magnitude bin thus, defines the empirical PMS locus (see \citet{2021MNRAS.504.2557D} for details). The Figure \ref{fig:age bin} ({\it Left}) shows the Hess plot of r$_{2}$ - i$_{2}$ versus r$_{2}$ CMD overplotted with the finally selected sources (red sources) and the empirical PMS locus (green solid curve) along with the 20 Myr Baraffe isochrone (black dashed curve). We also present the color distribution in each magnitude bin which defines the PMS locus in Figure \ref{fig:age bin} ({\it Right}).\\

\begin{figure}
	%\centering
	%\includegraphics[width=8.5cm, height=6cm]{i2Y_cmd_Yso.png}
	%\includegraphics[width=8.45cm, height=6cm]{age_hist.png}
	\includegraphics[scale=0.2]{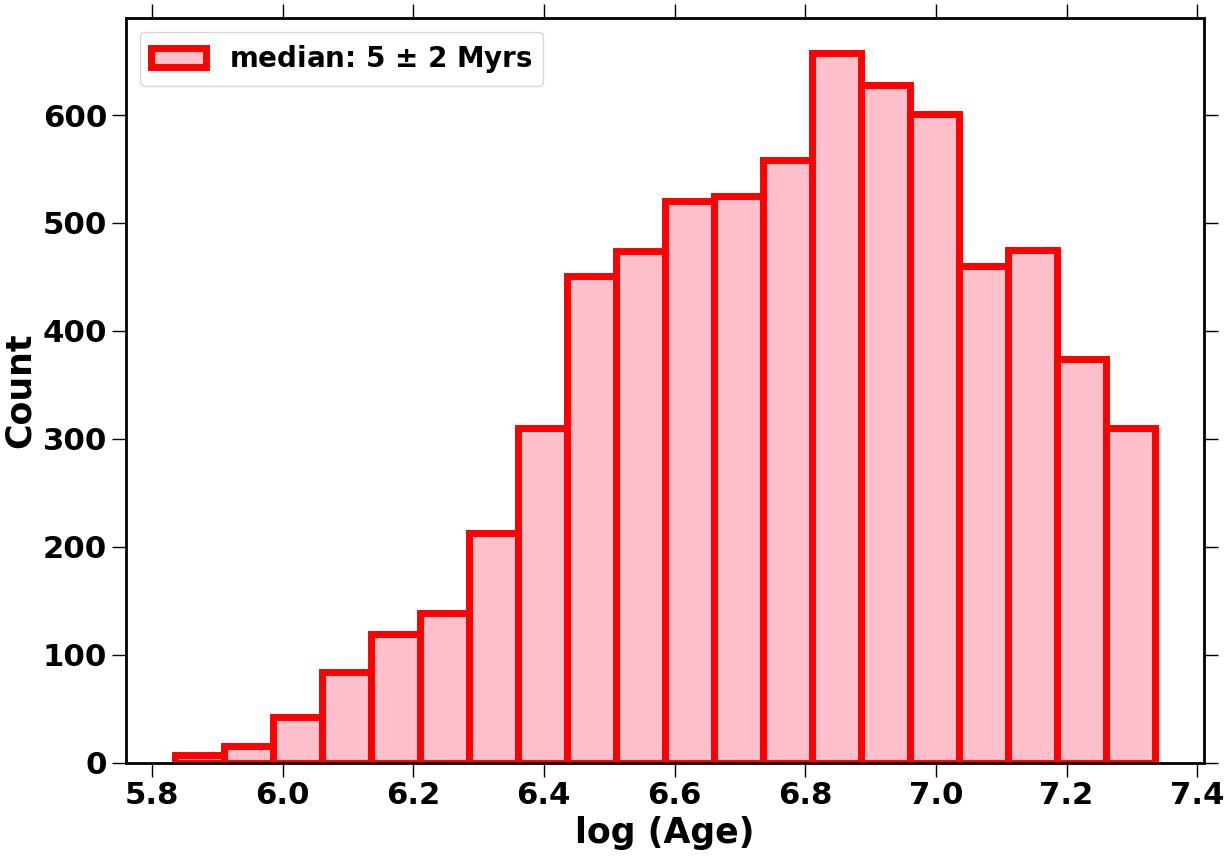}
	%\includegraphics[width=6.2cm, height=9cm]{ageBin_hist.png}
	%\includegraphics[scale = 0.38]{r2-Y_vs_r2.eps}%\scriptsize
	%\includegraphics[scale = 0.4]{z-K_vs_z.eps}%\scriptsize
	%\begin{small}  
	%\scriptsize
	%\linespread{0.8}
	\caption{ Histogram to represent the distribution of age among the selected sources (represented as red dots in Figure \ref{fig:age bin}). %The Blue histogram is overplotted to show the age distribution for sources within 2$\sigma$ limits of the mean age, after 5 iterations. Both the histograms are fitted with the Gaussian (solid black curves). The mean age for the entire unconverged distribution is 6.2 Myrs and marked by the yellow dashed line. the mean age for the entire 1$\sigma$ converged distribution is 5.3 Myrs and marked by the brown dashed line.
	 } 

%{\it Top left:}
% }
%\end{small}
\label{fig:age hist}
\end{figure}

We determine the age of these selected sources by fitting the Baraffe isochrones of various ages (available at an interval of log(t) = 0.01). %, since $>$ 95$\%$ of the above selected sources are located within the 20 Myr isochrone. 
The age is then assigned to each source based on its distance to the different isochrones. Since for any particular age, the available isochrones are a set of few discrete points (color and magnitude values), the age estimation based on the distance to these few points can be biased. Hence, we fit these discrete points using linear regression model with fifth order polynomial distribution to interpolate the isochrones. This interpolation generates a larger set of discrete points for any particular age and the accuracy of these predicted values (color and magnitude values) is $\ge$ 99$\%$ for all the isochrones of different ages. The interpolation of the isochrones thus, helps in improving the overall accuracy of this age estimation method. We then proceed to find, for each source, the two nearest isochrones with ages, say t$_{1}$ and t$_{2}$ and distances D$_{1}$ and D$_{2}$ respectively, from the source. The age is then calculated as the weighted average of the two ages t$_{1}$ and t$_{2}$. The inverse of the distances D$_{1}$ and D$_{2}$ are used as weights in order to calculate the weighted average (t) of the ages of the two isochrones as given in equation below:
\begin{center}
    $t = \frac{t_{2}D_{1} + t_{1}D_{2}}{D_{1} + D_{2}}$
\end{center}
%\begin{center}
%   $$\boldsymbol t = \frac{ \boldsymbol{t_{2}D_{1} + t_{1}D_{2}}}{ \boldsymbol{D_{1} + D_{2}}}$$
%\end{center}
%\end{center}
%Since the Baraffe isochrones are available at an interval of 1 Myr, we obtain a discrete age distribution among the sources. Hence, in order to obtain a continuous distribution, we simply divide their r$_{2}$ - i$_{2}$ color and r$_{2}$ mag into 0.1 mag bins and the mean age of the sources within is assigned to each bin. 
The weighted average t is thus, assigned as the age of the source. The process is repeated for all the selected sources. The median age of the field decontaminated sources within  18$^\prime$ is thus, obtained to be 6.5 $\pm$ 5 Myrs. We further converge this distribution to within 2 $\sigma$ limits from the mean age of the entire distribution after performing 8 iterations. The median age for the 2 $\sigma$ converged sample turns out to be 5 $\pm$ 2 Myrs. The Figure \ref{fig:age hist} shows histogram plot for the age distribution of the sources for the un-converged sources. Although for the above age calculation, we have reddened the Baraffe isochrones for an A$_{V}$ = 6 mag, we derive similar results (median age within 4 -- 6 Myrs) for an extinction variation between A$_{V}$ = 4.5 - 7.5 mag (\citealt{2010ApJ...713..871W, 10.1093/mnras/stv323}). This is expected because the reddening vector stays parallel to the isochrones for optical wavelengths. Hence, a variation in the extinction simply shifts the sources along the isochrones without thus, introducing any significant modification in the derived ages. Also, the derived age of the region remains within 4 - 6 Myrs for a distance variation ranging between $\sim$ 1500 - 1700 pcs (distance to Cygnus OB2 = 1600 $\pm$ 100 pcs (\citealt{10.1093/mnras/stz2548})). The other possible factors like binarity, optical variability, although add to the broadening of the color in CMDs of young star forming regions, however, may not affect the true age spread as well as the cluster parameters like IMF  significantly (\citealt{2017ApJ...836...98J, 2021MNRAS.504.2557D}).   %The age obtained by our analysis agrees quite well with that estimated by several other studies of the region. For example, \citet{2008MNRAS.386.1761D} analyse 200 A-type stars across the Cygnus OB2, using IPHAS photometry and find the age to be $\sim$ 5 Myrs. Similarly, \citet{10.1093/mnras/stv323} used a list of 169 massive OB stars to derive the age of the region as $\sim$ 4 - 5 Myrs using rotating stellar evolutionary models from \citet{2012A&A...537A.146E} while \citet{2010ApJ...713..871W} use X-ray sources to obtain 3.5 - 5.2 Myrs as the average age of the region. Recent studies by \citet{2018A&A...612A..50B, 2020A&A...644A..62C} perform spectroscopy of $\sim$ 60 OB-type stars (observed with INT, ISIS, OSIRIS instruments) and find that the age of the region ranges between 1 - 6 Myrs irrespective of the stellar model used for age estimation. We corroborate this result by verifying our age estimation with Parsec isochrone models (\citealt{2012MNRAS.427..127B}) in addition to the Baraffe models, for a mass range of 0.3 M$_{\odot}$ - 1.4 M$_{\odot}$ and derive the median age $\sim$ 4.5 $\pm$ 2 Myrs. 
The above analysis thus, confirms the median age of the central 18$^\prime$ region with that of $\le$ 10 Myrs as estimated by the previous studies (\citealt{2008MNRAS.386.1761D, 10.1093/mnras/stv323, 2018A&A...612A..50B}).

\subsection{Disk Fraction}
\label{sec: disk fraction}
Circumstellar disk evolution sets the timescale for planet formation and hence, measuring the disk fraction, that is, the fraction of stars surrounded by circumstellar disks for a certain cluster age, is an important parameter to give an insight into the star and planet formation in a young cluster (\citealt{2001ApJ...553L.153H, 2011ARA&A..49...67W, 2014prpl.conf..643H, 2014A&A...561A..54R}). %In general, the circumstellar disks dissipate within initial few million years of star formation with dissipation timescale longer for low mass stars than their massive counterparts (\citealt{2011ARA&A..49...67W, 2014MNRAS.442.2543Y, 2015A&A...576A..52R}).
Although in a young cluster, disk fraction depends upon various factors such as the metallicity, stellar density, environmental factors like external and internal photoevaporation (\citealt{2016AJ....151..115Y, 2021arXiv210411764Y, Thies_2010, 2016arXiv160501773G, 2019MNRAS.486.4354R}), a general trend of disk fraction declining with age is observed. It ranges between 60$\%$ - 80$\%$ for clusters like NGC 1333 (\citealt{2014A&A...561A..54R}), NGC 2023, RCW36 (\citealt{2018MNRAS.477.5191R}) with an age $<$ 1 Myr (e.g ) to 5$\%$ - 10$\%$ for clusters like LowCent-Crux (\citealt{2007ApJ...671.1784H}), 25 Orionis (\citealt{2016MNRAS.461..794P}) with age $\sim$ 10 Myrs. In this section we calculate the disk fraction for the central 18$^\prime$ region of Cygnus OB2. \\

\begin{figure}
	%\centering
	%\includegraphics[width=8.5cm, height=6cm]{i2Y_cmd_Yso.png}
	\includegraphics[scale=0.26]{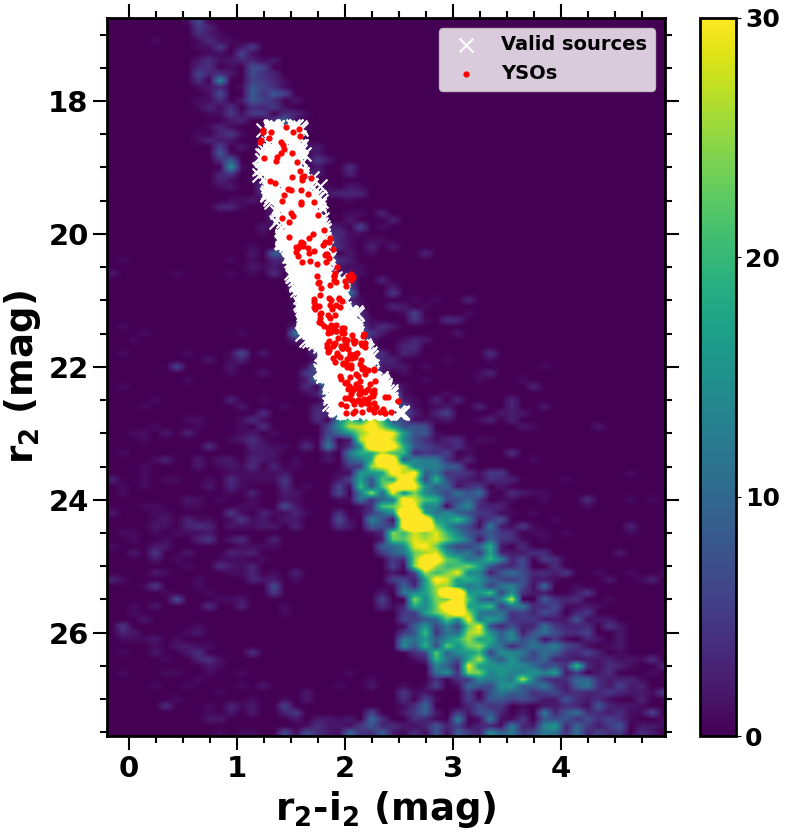}
	%\includegraphics[width=6.2cm, height=9cm]{ageBin_hist.png}
	%\includegraphics[scale = 0.38]{r2-Y_vs_r2.eps}%\scriptsize
	%\includegraphics[scale = 0.4]{z-K_vs_z.eps}%\scriptsize
	%\begin{small}  
	%\scriptsize
	%\linespread{0.8}
	\caption{ Hess plot of r$_{2}$ - i$_{2}$ versus r$_{2}$ CMD for the field subtracted sources. This Hess diagram is overplotted with the YSOs (Red circles) along with the sample selected to calculate the disk fraction (i.e the total number of sources) (White crosses).
	 } 

%{\it Top left:}
% }
%\end{small}
\label{fig:Disk fraction}
\end{figure}

In order to calculate the disk fraction, we consider the previously identified YSOs by \citet{Guarcello_2013} within the cluster field area of 18$^\prime$ radius. The previously identified YSOs are complete between 0.7 M$_\odot$ -- 2 M$_\odot$ (\citealt{Guarcello_2013}), which corresponds to 18.5 mag $\le$ r$_{2}$ $\le$ 22.5 mag at a distance $\sim$ 1600 pc and A$_{V}$ $\sim$ 6 mag. Hence, for estimating the disk fraction, we consider only those YSOs with optical counterparts within the mentioned r$_{2}$-band magnitude completeness range. The sample data used to calculate the disk fraction thus consists of only those field subtracted member sources which lie within 1 $\sigma$ limit of the pre-main sequence locus (Section $\ref{sec: age}$) and 18.5 mag $\le$ r$_{2}$ $\le$ 22.5 mag. Figure \ref{fig:Disk fraction} shows the Hess plot of r$_{2}$ - i$_{2}$ versus r$_{2}$ CMD for the field subtracted sources. This Hess diagram is overplotted with the YSOs (Red circles) along with the sample selected to calculate the disk fraction (i.e the total number of candidate members) (White crosses). We find that the ratio of the number of YSOs to that of the total number of sources, also termed as the disk fraction, turns out to be $\sim$ 9$\%$. This is however, a lower limit on the disk fraction as the previously identified YSOs are limited by the Spitzer IRAC Channel 2 sensitivity. This reason accounts for the lower disk fraction ($\sim$ 9$\%$) obtained by our analysis as compared to the 18$\%$ - 40$\%$ estimated by \citet{2016arXiv160501773G}. Cygnus OB2 has a lower disk fraction, in comparison to other young clusters like NGC 2264, CepOB3-East and West, which could be a result of external photoevaporation of circumstellar disks as a result of massive stars in vicinity. \\%The particularly low disk fraction in Cygnus OB2, as compared to other nearby clusters (\citealt{}) is presumed to be a consequence of feedback processes in the form of external photoevaporation of circumstellar disk in the region (\citealt{2016arXiv160501773G}). }}\\

%{\textbf{According to \citet{2016arXiv160501773G}, the disk fraction in the entire 1$^\circ$ $\times$ 1$^\circ$ is found to vary between 18$\%$ - 40$\%$ with the former value observed towards the central part of the cluster. In accord with our results, the recent study by \citet{2018MNRAS.477.5191R} with 69 MYStIX and SFiNCs young clusters reveals that the disk fraction drops to $\le$ 20$\%$ for a cluster age $\ge$ 4 Myrs. Similarly, when compared to nearby young clusters such as $\lambda$Ori, OriOB1b (\citealt{2010ApJ...722.1226H, 2020ApJ...893...56M}) within the similar age range i.e 4 -- 6 Myrs, the average disk fraction ranges between 15 -- 20$\%$ which is in excellent agreement with our results. However, in comparison to several other young clusters (within age $\sim$ 3 -- 4 Myrs) such as NGC 2264 (\citealt{2009AJ....138.1116S}), CepOB3b-East and West (\citealt{2012ApJ...750..125A}), AFGL 333/W3 (\citealt{Jose_2016}), IC348/U (\citealt{2018MNRAS.477.5191R}), NGC 2282 (\citealt{2015MNRAS.454.3597D}), where an average disk fraction of 30$\%$ - 50$\%$ is observed, the particularly lower disk fraction obtained above for the central region of Cygnus OB2 can be attributed to the feedback effect from the massive OB-type stars clustered towards the centre which induces external photoevaporation of the protoplanetary disks surrounding the low mass stars in vicinity (\citealt{2016arXiv160501773G})}}. \\ 

%\subsection{}
%\label{sec:maths} % used for referring to this section from elsewhere

%\begin{equation}
%    x=\frac{-b\pm\sqrt{b^2-4ac}}{2a}.
%	\label{eq:quadratic}
%\end{equation}

%Refer back to them as e.g. equation~(\ref{eq:quadratic}).

\section{Discussion}
\label{sec:discuss}

Rigorous studies of the low mass star formation in young massive Galactic clusters using multi-wavelength data sets are crucial to understand and solve some of the important yet unanswered questions such as the nature of IMF for stellar masses $<$ 0.5 M$_\odot$, the role of feedback driven cluster environment on the evolution of circumstellar disks, proportion of sub-stellar objects etc. The young massive association of Cygnus OB2 is a promising target for such purpose with its substantial massive as well as pre-main sequence population (\citealt{2007A&A...464..211A, 2009ApJS..184...84W}). %The ionization front driven by the massive stars towards the centre extends upto 50 pcs and has cleared the molecular gas from the surrounding cavity (\citealt{2006A&A...458..855S}). 
This paper presents the deepest and the widest optical photometry of Cygnus OB2 available as of yet. We detect a total of 713,529 sources with reliable data quality for objects detected down to the faint low mass end (Section \ref{sec:quality}). The preliminary data analysis performed with the deep HSC catalog suggests the presence of two sequences in various CMDs (Section \ref{sec:CMD}), the rightward sequence occupied by the PMS cluster members along with background contaminants. The previously identified YSOs overplotted on i$_{2}$-Y vs i$_{2}$ CMD in Figure \ref{fig:CMD} ({\it Left}) occupy the pre-main sequence branch in the CMD, mostly towards the right side of the isochrones of age $<$ 10 Myrs, as expected for a young association like Cygnus OB2 (e.g. \citealt{2017ApJ...836...98J,2021MNRAS.504.2557D}). We observe that the pre-main sequence segregation in various CMDs (Figure \ref{fig:CMD membership}) for the central region is consistent with most of the star formation being significantly clustered around the centre of this dynamically unevolved region (\citealt{2016MNRAS.460.2593W, 2020MNRAS.495.3474A}). The isochrone fitting done in Figure \ref{fig:CMD} {\it Left} suggests that $\sim$ $45\%$ of the total 713,529 sources detected in the region, lie within age less than 10 Myrs and a significant fraction of these sources ($\sim 12\%$) lie below the evolutionary track of mass less than 0.08 M$_\odot$. However, we caution the readers that this is an upper limit of candidate pre-main sequence population in the region as the estimated fraction is likely to be contaminated by the reddened background sources. More qualitative identification and classification of the YSOs in the entire HSC FoV of Cygnus OB2, both disk and diskless will be done in a future follow-up study using multi-wavelength photometry.\\
We perform the field decontamination of the central 18$^\prime$ region to get a statistical estimate of membership of the sources, using a control field located towards the periphery, which may be mostly contaminated with foreground and background stars. Approximately, 70$\%$ of the field decontaminated sources distinctly occupy the PMS branch with age less than 10 Myrs (Figure \ref{fig:field sub}). Since these statistically decontaminated members are used further to calculate age and disk fraction in the cluster field, we refine the membership with the help of an empirical PMS locus (see Section \ref{sec: age} for details). The median age of the central 18$^\prime$ region is $\sim$ 5 $\pm$ 2 Myrs. The age obtained by our analysis agrees quite well with that estimated by several other studies of the region. For example, \citet{2008MNRAS.386.1761D} analyse 200 A-type stars across the Cygnus OB2, using IPHAS photometry and find the age to be $\sim$ 5 Myrs. Similarly, \citet{10.1093/mnras/stv323} used a list of 169 massive OB stars to derive the age of the region as $\sim$ 4 - 5 Myrs using rotating stellar evolutionary models from \citet{2012A&A...537A.146E} while \citet{2010ApJ...713..871W} use X-ray sources to obtain 3.5 - 5.2 Myrs as the average age of the region. Recent studies by \citet{2018A&A...612A..50B, 2020A&A...644A..62C} perform spectroscopy of $\sim$ 60 OB-type stars (observed with INT, ISIS, OSIRIS instruments) and find that the age of the region ranges between 1 - 6 Myrs irrespective of the stellar model used for age estimation. We corroborate this result by verifying our age estimation with Parsec isochrone models (\citealt{2012MNRAS.427..127B}) in addition to the Baraffe models, for a mass range of 0.3 M$_{\odot}$ - 1.4 M$_{\odot}$ and derive the median age $\sim$ 4.5 $\pm$ 2 Myrs. Cygnus OB2 is a part of the larger Cygnus X giant molecular cloud which formed approximately 40 - 50 Myrs ago. The star formation towards Cygnus OB2 region however, has mainly taken place in the last 10 - 20 Myrs with the last star formation activity peaking around 3 - 5 Myrs ago (\citealt{2008hsf1.book...36R, 2012A&A...543A.101C, Comern2016RedSA, 2018A&A...612A..50B, 2020A&A...644A..62C}). This may suggest the substantial pre-main sequence population with the median age $\sim$ 5 Myrs in the region as obtained with our data analysis.\\

We obtain a disk fraction of $\sim$ 9$\%$ for this cluster field using the already known YSOs in the region. There is a wide variety of disk fractions measured in young clusters. An average disk fraction of 30$\%$ - 50$\%$ is observed in several young clusters (within age $\sim$ 3 -- 6 Myrs) such as NGC 2264 (\citealt{2009AJ....138.1116S}), CepOB3b-East and West (\citealt{2012ApJ...750..125A}), AFGL 333/W3 (\citealt{Jose_2016}), IC348/U (\citealt{2018MNRAS.477.5191R}) and NGC 2282 (\citealt{2015MNRAS.454.3597D}). However, recent studies of some nearby young clusters (\citealt{2010ApJ...722.1226H, 2016arXiv160501773G, 2018MNRAS.477.5191R}) show considerably smaller disk fractions. For example, the recent study by \citet{2018MNRAS.477.5191R} with 69 MYStIX and SFiNCs young clusters reveals that the disk fraction could drop to values $\le$ 15$\%$ for a cluster age $\ge$ 4 Myrs, which is consistent with our results. The particularly low disk fraction obtained for the central region of Cygnus OB2 and such other clusters which lie at the lower end of the spectrum of disk fractions, may be attributed to either the evolutionary effect or the feedback effect from the massive OB-type stars in vicinity (\citealt{2016arXiv160501773G}). In this work we cannot conclusively pinpoint the exact reason, however, evolutionary effects or external photo-evaporation could be some of the possible reasons for the observed low disk fractions.\\

% The low disk fraction is presumed to be a consequence of feedback processes in the form of external photoevaporation of circumstellar disks.The obtained disk fraction is however a lower limit on the disk fraction as the previously identified YSOs are limited by the Spitzer IRAC Channel 2 sensitivity.}} \\% We plan to perform a deeper YSO analysis for the entire observed region and hence, an elaborate disk fraction analysis, using deep multi-wavelength data including the presented HSC data as a part of further work. }}\\
 %Another interesting prospect is to investigate if there has been any event of triggering of star formation in the region owing to the high energy radiative feedback generated by the central massive stars or due to the jets/outflow activity from the YSOs present in the region.
%The deep optical photometry (r$_2$ $\sim 28 mag$) attained with the HSC data presented here will also be useful to detect sub-stellar objects down to $\le 0.07 M_\odot$ i.e. the Lithium-burning limit. 
The significant census of low mass and sub-stellar sources detected with deep HSC photometry (r$_2$ $\sim$ 28 mag) will serve as an excellent statistical sample for further studies to test the effect of feedback driven environmental conditions of Cygnus OB2 on low mass population across the region. %circumstellar disk evolution (e.g \citealt{2016arXiv160501773G, 2019MNRAS.485.1489W}), sub-stellar population and IMF across the region. 
To conclude, we find from our preliminary analysis that in accordance with the literature, Cygnus OB2 is a young active star-forming region (age $<$ 10 Myr) with a substantial pre-main sequence population. The deep multi-wavelength studies are essential to understand low mass star formation in the region and will be the area of focus in our future works. \\

\section{Summary and Future Works}
\label{sec: sumup}

This paper presents the deepest and the widest optical catalog of the young feedback-driven OB association of Cygnus OB2.\\%, which is among the most massive Galactic OB-associations with a rich young population.\\

1) A 1.5$^\circ$ diameter area of Cygnus OB2 was observed with Subaru Hyper Suprime-Cam (HSC) in 4 filters namely r$_{2}$, i$_{2}$, z and Y. The observations were taken in excellent seeing conditions ranging between 0.5$^{\prime\prime}$--0.7$^{\prime\prime}$. The observed raw data was reduced using HSC pipeline version 6.7.\\

2) The final HSC catalog contains only those point sources which have at least 2-band detection and additionally, have internal astrometric error $\le$ 0.1$^{\prime\prime}$ along with photometric error $\le$ 0.1 mag in individual bands. A total of 713,529 sources are detected with 699,798 sources having a must detection in Y-band, 685,511 sources in z-band, 622,011 in i$_{2}$ and 358,372 sources in r$_{2}$-band.\\ %The detected sources exhibit a magnitude offset of 0.03 mag, 0.01 mag, 0.01 mag, 0.01 mag in Y, z, $i_{2}$ and $r_{2}$-band respectively with that of Pan-STARRS photometry and within an error of $<= 0.1 mag$ in individual bands.\\

3) We detect sources down to 28.0 mag, 27.0 mag, 25.5 mag and 24.5 mag in r$_{2}$, i$_{2}$, z and Y-band respectively. Coupled with a distance of 1600 pc for an age ranging between 5 $\pm$ 2 Myrs and extinction A$_V$ $\sim$ 6 -- 8 mag, we achieve $\sim$ 90\% completeness down to a stellar mass $\sim$ 0.03 -- 0.06 M$_\odot$ and $\sim$ 0.03 -- 0.04 M$_\odot$ i.e $<$ Lithium burning limit, in i$_{2}$ and z-band respectively. The corresponding mass completeness limit is down to $\sim$ 0.02-0.03 M$_\odot$ and $\sim$ 0.15-0.30 M$_\odot$ in Y and r$_{2}$-bands, respectively.\\% Further analysis of the color-magnitude diagrams fitted with Baraffe isochrones suggest an age of $<$ 10 Myr for the region with a significant number of candidate pre-main sequence sources. \\

4) %The preliminary data analysis suggests the presence of a distinct PMS branch. 
The median age of the central region of Cygnus OB2 ranges between 4 -- 6 Myrs for an A$_{V}$ ranging between 4.5 -- 7.5 mag and distance between 1500 -- 1700 pcs. We obtain a disk fraction $\sim$ 9$\%$ in the central cluster, which is however a lower limit given the restricted completeness of the already known YSOs.\\

As the next step, we plan to adopt a multi-wavelength approach by combining the presented HSC optical data with other existing data from UKIDSS, 2MASS and Spitzer surveys to carry out a detailed analysis of the YSOs present in the region. %We aim to perform a detailed disk fraction analysis based on the newly obtained YSOs and the role of externally induced photoevaporation on it, which would help to understand the evolution of protoplanetary disks better and how it gets affected by the cluster environment. 
In addition to this we would use our deep optical photometry presented in this paper, coupled with other data sets to evaluate cluster parameters like IMF for very low mass stars ($<$ 0.1 M$_\odot$) along with identification and characterization of sub-stellar objects like brown dwarfs and understand the role of feedback-driven environment of Cygnus OB2 on such parameters.\\

\section{Data Availability}
A sample table of the HSC catalog is presented in Table \ref{tab:sample data}. The complete catalog is provided as online material.\\

\begin{table*}
	\centering
	\begin{threeparttable}
	\caption{ Sample table of HSC catalog data. The complete table is available as online material. }
	\label{tab:sample data}
	\begin{tabular}{ |l|l|l|l|l|l|l|l|l|l|l| } % four columns, alignment for each
		\hline
		Source & RA & Dec & r$_{2}$ & r$_{2_{err}}$ & i$_{2}$ & i$_{2_{err}}$ & z & z$_{err}$ & Y & Y$_{err}$ \\
		    & (deg) & (deg) & (mag) & (mag) & (mag) & (mag) & (mag) & (mag) & (mag) & (mag) \\
		\hline
		1 & 308.69298 & 41.86609 & 25.728 & 0.019 & 23.568 & 0.006 & 22.090 & 0.008 & 21.434 & 0.008 \\
		2 & 308.83647 & 41.86581 & 24.790 & 0.010 & 22.666 & 0.003 & 21.175 & 0.004 & 20.515 & 0.004 \\
		3 & 308.70283 & 41.86674 & 26.425 & 0.044 & 24.641 & 0.018 & 22.859 & 0.015 & 22.154 & 0.016 \\
		4 & 308.84554 & 41.86651 & 25.894 & 0.028 & 22.267 & 0.002 & 20.231 & 0.002 & 19.183 & 0.001 \\
		5 & 308.79625 & 41.86680 & 24.398 & 0.007 & 22.314 & 0.002 & 21.279 & 0.005 & 20.026 & 0.002 \\
		\hline
	%\caption{Table 1 shows various parameters obtained by analysis of final HSC catalog in individual filters. Here}
	%\label{tab:HSC Analysis}
	\end{tabular}
	%\begin{tablenotes}\footnotesize
%	\item [a] Magnitude of the brightest object detected
%	\item [b] Magnitude of the faintest object detected
%	\item [c] Magnitudes rounded off to 0.2 mag
%	\item [d] Mass corresponding to magnitude with 90\% completeness for A$_v: 6 mag-8 mag$ and age: 3-5 Myrs.
	%\end{tablenotes}
	\end{threeparttable}
\end{table*}

\section*{Acknowledgements}
The authors thank the referee for the useful constructive comments which has refined the overall structure and quality of this paper. This research is based on data collected at Subaru Telescope with Hyper Suprime-Cam, which is operated by the National Astronomical Observatory of Japan. We are honored and grateful for the opportunity of observing the Universe from Mauna Kea, which has the cultural, historical and natural significance in Hawaii. We are gateful to The East Asian Observatory which is supported by The National Astronomical Observatory of Japan; Academia Sinica Institute of Astronomy and Astrophysics; the Korea Astronomy and Space Science Institute; the Operation, Maintenance and Upgrading Fund for Astronomical Telescopes and Facility Instruments, budgeted from the Ministry of Finance (MOF) of China and administrated by the Chinese Academy of Sciences (CAS), as well as the National Key R\&D Program of China (No. 2017YFA0402700). The authors thank the entire HSC staff and HSC helpdesk for their help. We would like to thank S.Mineo, H.Furusawa, Y.Yamada and M.Kubo in HSC helpdesk team for useful discussions regarding the data reduction. We thank NAOJ for providing access to hanaco account which was used to perform some initial stages of data reduction. We gratefully acknowledge the use of high performance computing facilities at IUCAA, Pune for the HSC data reduction. We thank I.Baraffe for providing us with isochrone models for an interval of log (Age) = 0.01, through personal communication. We use Pan-STARRS and GAIA ED3 data for data quality checks. The Pan-STARRS1 Surveys (PS1) and the PS1 public science archive have been made possible through contributions by the Institute for Astronomy, the University of Hawaii, the Pan-STARRS Project Office, the Max-Planck Society and its participating institutes, the Max Planck Institute for Astronomy, Heidelberg and the Max Planck Institute for Extraterrestrial Physics, Garching, The Johns Hopkins University, Durham University, the University of Edinburgh, the Queen's University Belfast, the Harvard-Smithsonian Center for Astrophysics, the Las Cumbres Observatory Global Telescope Network Incorporated, the National Central University of Taiwan, the Space Telescope Science Institute, the National Aeronautics and Space Administration under Grant No. NNX08AR22G issued through the Planetary Science Division of the NASA Science Mission Directorate, the National Science Foundation Grant No. AST-1238877, the University of Maryland, Eotvos Lorand University (ELTE), the Los Alamos National Laboratory, and the Gordon and Betty Moore Foundation. This work has made use of data from the European Space Agency (ESA) mission GAIA processed by Gaia Data processing and Analysis Consortium (DPAC: https://www.cosmos.esa.int/web/gaia/dpac/consortium). PP and JJ acknowledge the DST-SERB, Gov. of India for the start up research grant (No: SRG/2019/000664). \\

%%%%%%%%%%%%%%%%%%%%%%%%%%%%%%%%%%%%%%%%%%%%%%%%%%



\bibliographystyle{mnras}
\bibliography{example} % if your bibtex file is called example.bib

%\begin{thebibliography}{99}
%\end{thebibliography}

%%%%%%%%%%%%%%%%%%%%%%%%%%%%%%%%%%%%%%%%%%%%%%%%%%

%%%%%%%%%%%%%%%%% APPENDICES %%%%%%%%%%%%%%%%%%%%%

\appendix

\section{}
\label{sec:proplyds}
\begin{figure*}
	%\centering
	\includegraphics[scale=0.37]{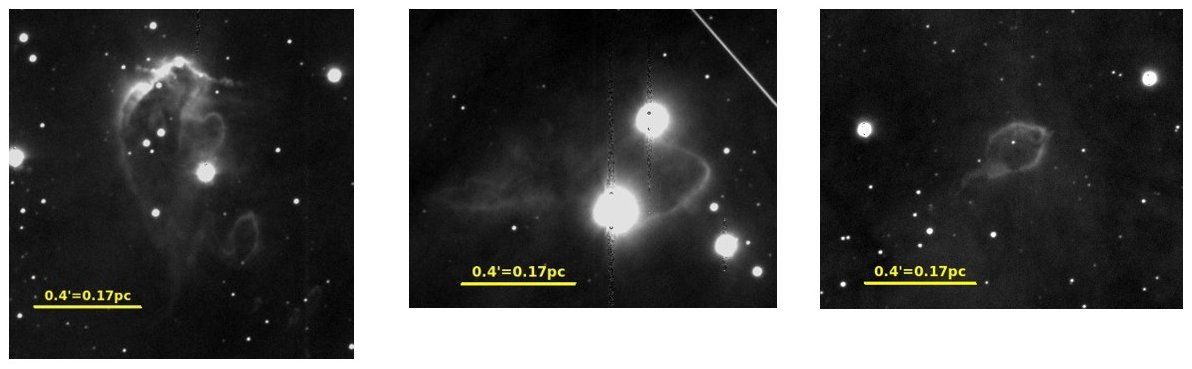}
	%\includegraphics[scale = 0.2]{cyg_centre1.eps}
	%\includegraphics[scale = 0.38]{r2-Y_vs_r2.eps}%\scriptsize
	%\includegraphics[scale = 0.4]{z-K_vs_z.eps}%\scriptsize
	%\begin{small}  
	%\scriptsize
	%\linespread{0.8}
	\caption{ Images of proplyds/globules/globulettes in Cygnus OB2 in $r_{2}$-band with their central co-ordinates {\it Upper Left} RA: 20:34:46.28; Dec: +40:52:36.9 {\it Upper Right} RA: 20:34:14.4438; Dec: +41:07:39.961 {\it Bottom Left} RA: 20:33:12; Dec: +40:41:48.657 {\it Bottom Middle} RA: 20:34:47; Dec: +41:14:45 {\it Bottom Right} RA: 20:34:53.6; Dec: +40:48:14.
	 }
%{\it Top left:}
% }
%\end{small}
\label{fig:Proplyds}
\end{figure*}

We present in Figure \ref{fig:Proplyds}, interesting images from HSC-$r_{2}$-band for a few  proplyds/globules/globulettes identified by \citealt{2012ApJ...746L..21W}, with centre of the regions mentioned in the sub-caption below.\\

\begin{figure*}
	%\centering
%	\includegraphics[width=19cm, height=14cm]{proplyds1.png}
	\includegraphics[scale=0.38]{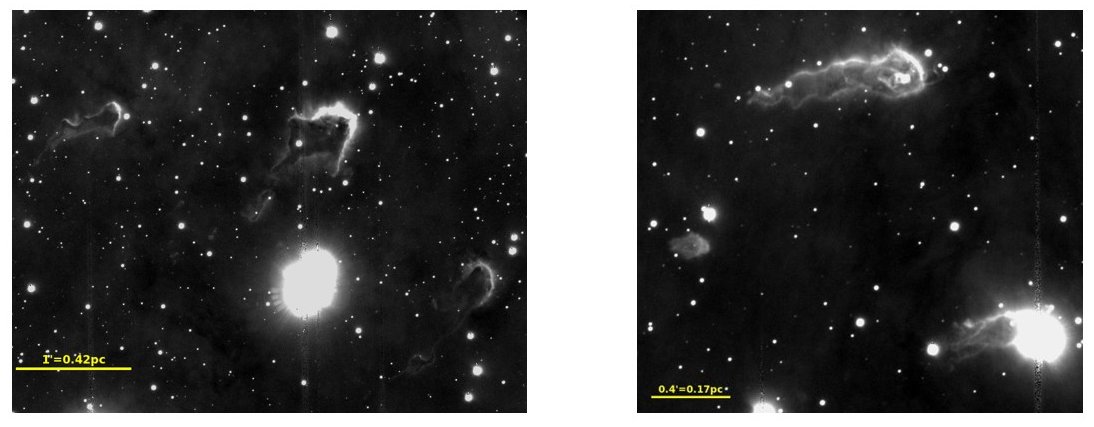}
	%\includegraphics[scale = 0.4]{z-K_vs_z.eps}%\scriptsize
	%\begin{small}  
	%\scriptsize
	%\linespread{0.8}
%	\caption{ Images of proplyds/globules/globulettes in Cygnus OB2 in $r_{2}-band$ with their central co-ordinates {\it Upper Left} RA: 20:34:43.3; Dec: +40:53:13.7 {\it Upper middle} RA: 20:33:12; Dec: +40:41:48.657 {\it Upper Right} RA: 20:34:36.4; Dec: +40:51:54.7 {\it Bottom Left} RA: 20:34:47; Dec: +41:14:45 {\it Bottom Right} RA: 20:34:10.5; Dec: +41:06:59
%	 }
	%\caption{ Images of proplyds/globules/globulettes in Cygnus OB2 in $r_{2}$-band with their central co-ordinates {\it Left} RA: 20:34:46.28; Dec: +40:52:36.9 {\it Right} RA: 20:34:14.4438; Dec: +41:07:39.961
	% }
	
%{\it Top left:}
% }
%\end{small}
%\label{fig:proplyd1}
\end{figure*}

%\section{}
%\subsection{Input count plots}
%We present in Figure \ref{fig:CountInput} the spatial plots for input count values in individual HSC filters.\footnote{** Should these spatial plots be included?} \\

%\begin{figure*}[H]
	%\centering
%	\includegraphics[width=8.6cm, height=6.5cm]{countInput_R2_spatial.png}
%	\includegraphics[width=8.6cm, height=6.5cm]{countInput_I2_spatial.png}
	%\includegraphics[scale = 0.2]{cyg_centre1.eps}
	%\includegraphics[scale = 0.38]{r2-Y_vs_r2.eps}%\scriptsize
	%\includegraphics[scale = 0.4]{z-K_vs_z.eps}%\scriptsize
	%\begin{small}  
	%\scriptsize
	%\linespread{0.8}
	%\caption{ Scatter plots of HSC magnitudes {\it versus} Error in individual filters. All the sources have error $<=$ 0.1.
	% }
%{\it Top left:}
% }
%\end{small}
%\label{fig}
%\end{figure*}

%\begin{figure*}[H]
	%\centering
%	\includegraphics[width=8.6cm, height=6.5cm]{countInput_Z_spatial.png}
%	\includegraphics[width=8.6cm, height=6.5cm]{countInput_Y_spatial.png}
	%\includegraphics[scale = 0.2]{cyg_centre1.eps}
	%\includegraphics[scale = 0.38]{r2-Y_vs_r2.eps}%\scriptsize
	%\includegraphics[scale = 0.4]{z-K_vs_z.eps}%\scriptsize
	%\begin{small}  
	%\scriptsize
	%\linespread{0.8}
%	\caption{ Spatial plots for input count values in HSC-$r_{2}$ ({\it Top Left}), $i_{2}$ ({\it Top Right}), z ({\it Bottom Left}) and Y-band (({\it Bottom Right})) filters.
%	 }
%{\it Top left:}
% }
%\end{small}
%\label{fig:CountInput}
%\end{figure*}

\section{Transformation Equations}
\label{sec:transform eq}
The transformation equations\footnote{During the data reduction the coefficients used by the pipeline are as mentioned in the transformation equations} used to convert magnitudes from Pan-STARRS system to Subaru HSC system in individual bands in order to plot the magnitude offsets are given below:

\begin{equation}
    \begin{split}
    Y_{HSC} & = Y_{Pan-STARRS} - 0.001952 + (0.19957 (Y-z)_{Pan-STARRS}) \\
    & + (0.216821 ((Y-z)^2)_{Pan-STARRS})
    \end{split}
\end{equation}
\begin{equation}
    \begin{split}
    z_{HSC} & = z_{Pan-STARRS} - 0.005585 - (0.220704 (z-Y)_{Pan-STARRS}) \\
    & - (0.298211 ((z-Y))^2_{Pan-STARRS})
    \end{split}
\end{equation}
\begin{equation}
    \begin{split}
    i_{2_{HSC}} & = i_{2_{Pan-STARRS}} + 0.001653 - (0.206313 (i_{2}-z)_{Pan-STARRS} ) \\
    & - (0.016085 (i_{2}-z)^2_{Pan-STARRS}) 
    \end{split}
\end{equation}	
\begin{equation}
    \begin{split}
    r_{2_{HSC}} & = r_{2_{Pan-STARRS}} + 0.000118 - (0.00279 (r_{2}-i_{2})_{Pan-STARRS} ) \\
    & - (0.014363 (r_{2}-i_{2})^2_{Pan-STARRS}) \\
    \end{split}
\end{equation}

The reddening laws (\citealt{Wang_2019}) adopted by us to correct the Baraffe isochrones for extinction in the Pan-STARRS are mentioned below: \\

\begin{center}
$\frac{A_r}{A_V} = 0.843\pm0.006 $ \\
$\frac{A_i}{A_V} = 0.628\pm0.004 $ \\
$\frac{A_z}{A_V} = 0.487\pm0.003 $ \\
$\frac{A_y}{A_V} = 0.395\pm0.003 $ \\
\end{center}

These equations were used to convert the absolute Pan-STARRS magnitudes to apparent magnitudes using distance = 1600 parseccs and A$_V$ = 6 mag. The transformation equations mentioned above are then used to convert to HSC photometric system to redden the isochrones appropriately.

\section{}
\label{sec: spatial astrometry}
We present here the spatial distribution of astrometric offset of HSC data with respect to Pan-STARRS DR1 and Gaia EDR3 data.
% Don't change these lines

\begin{figure}
	%\centering
	\includegraphics[scale=0.169]{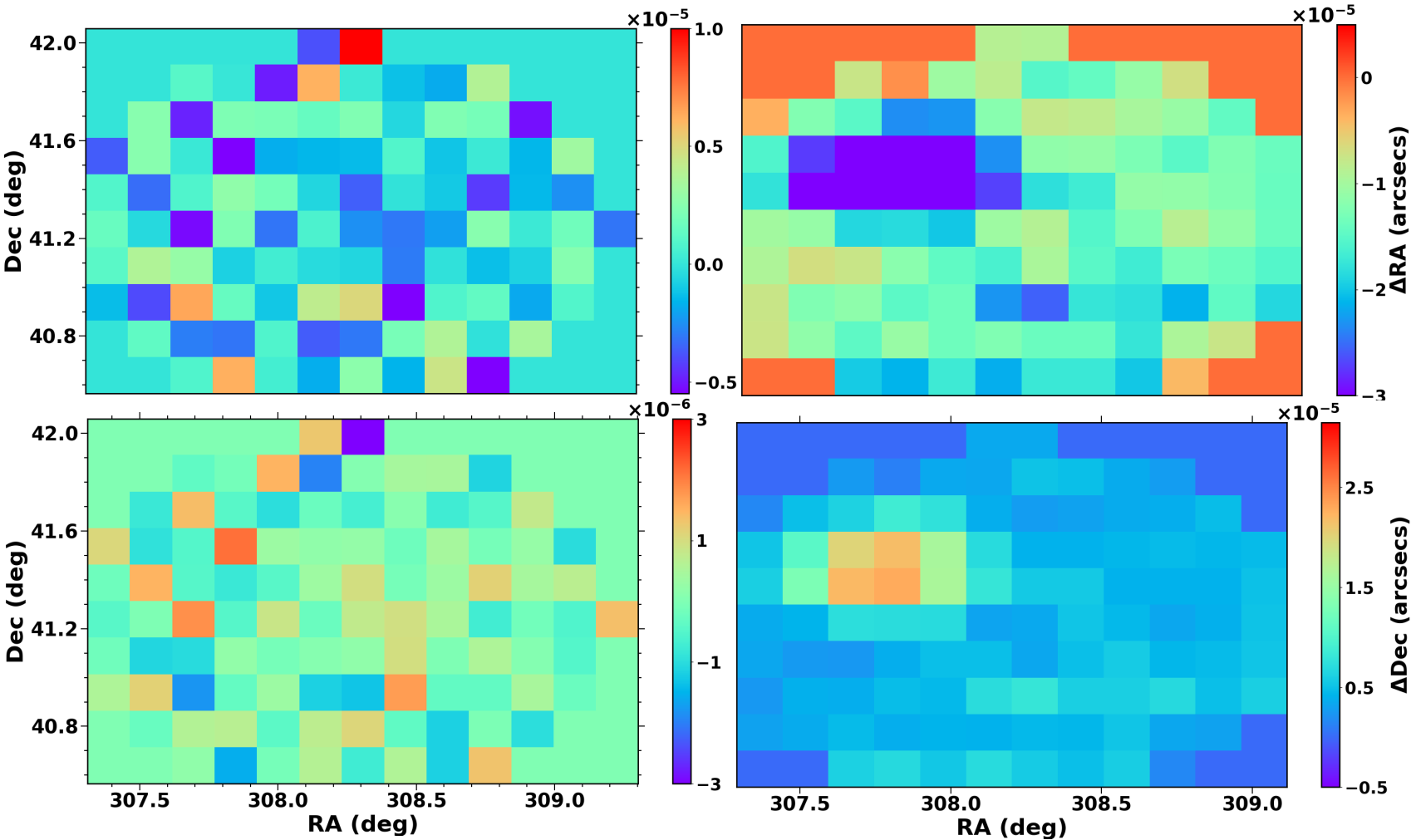}
	%\includegraphics[scale=0.14]{spatial_Dec_HSCxPS.png}
	%\includegraphics[scale=0.18]{Del_RA_spatial.png}
	%\includegraphics[scale=0.18]{Del_Dec_spatial.png}
	%\includegraphics[scale = 0.2]{cyg_centre1.eps}
	%\includegraphics[scale = 0.38]{r2-Y_vs_r2.eps}%\scriptsize
	%\includegraphics[scale = 0.4]{z-K_vs_z.eps}%\scriptsize
	%\begin{small}  
	%\scriptsize
	%\linespread{0.8}
	\caption{ Spatial plots signifying the variation of astrometric offset in Right Ascension ({\it Upper Left}) and Declination ({\it Bottom Left}) between HSC and Pan-STARRS data as well as HSC and Gaia EDR3 data ({\it Upper Right} and {\it Bottom Right}) across the entire region. The spatial maps are obtained by binning the RA and Dec parameter space into 10$^{\prime}$ $\times$ 10$^{\prime}$ bins across the entire observed region. The colorbar indicates the mean uncertainity in RA ({\it Left}) and Dec ({\it Right}) of each bin. %The observed internal astrometric error ranges between 0.01$^{\prime\prime}$ - 0.03$^{\prime\prime}$, with almost uniform distribution throughout the region.
	 }%{\it Top left:}
% }
%\end{small}
\label{fig:external astro spatial}
\end{figure}

%%%%%%%%%%%%%%%%%%%%%%%%%%%%%%%%%%%%%%%%%%%%%%%%%%

% Don't change these lines
\bsp	% typesetting comment
\label{lastpage}
\end{document}